\documentclass[12pt, a4paper, oneside, openright]{book} 
\usepackage[T1]{fontenc}
\usepackage{url} 
\usepackage{graphicx}
\usepackage{mathrsfs}
\usepackage{mathtools}
\usepackage{titlesec}
\usepackage[titles]{tocloft}
\usepackage{makeidx}
\usepackage{tikz}
\usepackage{setspace}
\usepackage{hyperref}

\titleformat{\chapter}[frame]{\normalfont}{\filright{ }CHAPTER \thechapter { }}{8pt}{\LARGE\rm\bf\filcenter}

\titleformat{\section} {\titlerule
\vspace{-4ex}%
\normalfont\sc} {\thesection.}{.5em}{}

\titleformat{\subsection} {\titlerule
\vspace{-4ex}%
\normalfont\sl} {\thesubsection.}{.5em}{}

\titleformat{\abstract}[frame]{\normalfont}{\center Abstract}{8pt}{\Huge\rm\filcenter}

\titlespacing*{\section}
{0pc}{*7}{*2}
\titlespacing*{\subsection}
{0pc}{*6}{*2}

\makeindex

\def\d{{\mathrm{d}}}

\def\tr{{\mathrm{tr}}}
\def\diag{{\mathrm{diag}}}
\def\ln{{\mathrm{ln}}}

\begin{document}

\newenvironment{warning}{{\noindent\bf Warning: }}{\hfill $\eof$\break}

\frontmatter

\begin{titlepage}

\begin{center}
  \parbox{100mm}{\filcenter\LARGE\textsc{\textbf{Applications of,\\ and Extensions to,\\Selected Exact Solutions\\ in General Relativity}}}\par
  \vspace*{\stretch{2}}
  \parbox{100mm}{\center by}\par
    \vspace*{\stretch{0.2}}
  \parbox{100mm}{\center\Large\sc{Bethan Cropp}}\par
     \vspace*{\stretch{1}}
  \parbox{100mm}{\center
                         A thesis\\ 
                         submitted to Victoria University of Wellington\\ 
                         in fulfilment of the\\
                         requirements for the degree of\\  
                         Master of Science \\ 
                         in Mathematics.}\par
 \parbox{100mm}{\center Victoria University of Wellington\\ \number\year}

 \end{center}

\pagestyle{empty}
\if@twoside\cleardoublepage\else\clearpage\fi
\mbox{}\vspace{20mm}
\begin{center}
{\Large\textbf{\textsc{Abstract}}}
\end{center}
In this thesis we consider several aspects of general relativity relating to exact solutions of the Einstein equations. In the first part gravitational plane waves in the Rosen form are investigated, and we develop a formalism for writing down any arbitrary polarisation in this form. In addition to this we have extended this algorithm to an arbitrary number of dimensions, and have written down an explicit solution for a circularly polarized Rosen wave. In the second part a particular, ultra-local limit along an arbitrary timelike geodesic in any spacetime is constructed, in close analogy with the well-known lightlike Penrose limit. This limit results in a Bianchi type I spacetime. The properties of these spacetimes are examined in the context of this limit, including the Einstein equations, stress-energy conservation and Raychaudhuri equation. Furthermore the conditions for the Bianchi type I spacetime to be diagonal are explicitly set forward, and the effect of the limit on the matter content of a spacetime are examined.
\end{titlepage}

\chapter*{Acknowledgments}

First and foremost, I wish to thank my supervisor, Matt Visser. His support has been vital to this undertaking. The insight and patience he has shown have been unparalleled. In particular I would also like to specifically thank him for his generosity in sending me to the GR19 and ERE2010 conferences, which I feel I have gained much from. 

Throughout this year I have enjoyed the company of the other applied mathematics students, Jozef Skakala, Gabriel Abreu, Nicole Walters and more recently Valentina Baccetti and Kyle Tate. 

Finally I would like to acknowledge the various people, too numerous to list, within the mathematics and physics departments of Victoria who have made the last five years as enjoyable and educational as they were.

\addcontentsline{toc}{chapter}{Acknowlegdements}

\chapter{Notation}

\begin{itemize}

\item Unless otherwise specified $c=1$, $G_N=1$.

\item We work with signature $(-1, 1, 1, 1)$.

\item Throughout $\nabla^2$ is used to denote the four dimensional Laplacian, with ${}^{(3)}\nabla^2$ the three dimensional Laplacian. Likewise when clarification is needed the dimension will be denoted in the same manner, e.g. ${}^{(3)}R$.

\end{itemize}

\noindent
Throughout this thesis we consider tensors in a number of different dimensions. To help clarify which number of dimensions we are working with in general:
\begin{itemize}
\item lower-case letters from the start of the alphabet $\left\{a, b, c, d\right\}$ will be used for indices over four dimensions.
\item lower-case letters from the middle of the alphabet $\left\{i, j, k, l \right\}$ will indicate indices over three dimensions.
\item upper-case letters from the start of the alphabet $\left\{A, B, C, D\right\}$ will denote indices over two dimensions.
\end{itemize}

\noindent
Further we will use $\left\{\alpha, \beta, \gamma, \delta \right\}$ for labelling of an orthonormal basis, and we leave it for the reader to determine the dimension based on the tensor indices used.

\chapter{Preface}

This thesis focuses on two separate topics both of which are related to physically interesting exact solutions in classical general relativity. 

\vspace{15pt}

\noindent
The first research problem concerns polarisations of strong-field gravitational plane waves. This is one case where an exact solution is simple enough to be mathematically tractable but, although highly idealised, may still capture aspects that are of physical relevance. 

As with many well-studied metrics in general relativity, more than one way has been discovered to represent these gravitational waves, known respectively as the Rosen and the Brinkmann forms. These two forms have their respective advantages, depending on the scenario considered, as would be expected. However it is important to know, in theory, how to construct solutions for each form.  When there are two ways of representing the same phenomena, and it appears much harder, or impossible, to formulate solutions in one form than the other, then it may indicate that work in this respect is incomplete.

In the case of gravitational waves, it was trivial to represent arbitrary polarisations in the Brinkmann form, while it was unclear how to construct arbitrary polarisations in the Rosen form. As such we investigated: 

\begin{itemize}

\item{Is there a clear way to construct arbitrary polarisations in the Rosen form? In particular is there a simple algorithmic construction to do this?}

\item{Is it possible to write down particularly useful polarisations, such as circular polarisation in the Rosen form?}

\end{itemize}
These questions have been answered positively, and a general method for constructing such polarisation modes has been developed. 

\vspace{15pt}

\noindent
One context in which plane waves, and their generalisations, {\it pp} waves, are studied is as the result of a Penrose limit. This well-known limit approximates the spacetime along any arbitrary null geodesic in any spacetime. We attempted to find a similar, timelike, limit. As such the questions investigated in this thesis are:

\begin{itemize}

\item{Is it possible to develop a similar limit for timelike geodesics, and what is the result of this limit?}

\item{In what way does this limit share the properties of the Penrose limit?}

\item{What happens to any matter content when taking the limit?}

\end{itemize}

We have developed a limit that appears to be a close timelike analogue to the Penrose limit, although our construction does not share all of the important properties of Penrose limits. It is also found to be a type of ultra-local limit, whereby the evolution of neighbouring points decouples.

The result of this limit was a Bianchi type I spacetime. This is another well-studied exact solution which is simple enough to be mathematically tractable. It has most frequently been studied as a basic cosmological model. This has led to most studies making context-appropriate assumptions that in the context of this thesis should not be made. In particular most studies have worked with a diagonal metric, and mostly examined vacuum, perfect fluid or dust scenarios. In counterpoint, in our context we found the most general scenario for the stress-energy was that the limit would erase information about the flux, but without more stringent conditions for general spacetimes. Therefore some further questions were raised:

\begin{itemize}

\item{What can we say about Bianchi type I spacetimes whilst keeping full generality?}

\item{What can we say about when it is appropriate to have a diagonal metric?}

\end{itemize}

We have provided a set of necessary and sufficient conditions for the metric to be diagonal. We have furthermore made a number of calculations relating to the Einstein equations, covariant conservation and the Raychaudhuri equation in general Bianchi type I spacetimes, simplifying as much as possible, and interpreting the results.

\section*{Arrangement of this thesis}

This thesis has been arranged into four main chapters plus conclusions. The first is a general introduction to some aspects of general relativity, many of which are used later in the research sections of the thesis. Much of the information in this chapter could be found in a good introductory textbook in general relativity. 

The second chapter introduces and reviews the topic of strong-field gravitational waves before going on to examine the problem of constructing strong-field polarisations of the Rosen form. From section (\ref{pp:arbit}) onwards the information presented is primarily original research, results from which have been presented in \cite{Cropp1, Cropp3}.

The third chapter forms another background section providing information on some topics of particular relevance to chapter four, which develops and explores a Penrose-like limit. Unlike the first chapter, much of this information (with the exception of the discussion of the FLRW metric) would not be included in a basic textbook, but is more advanced information familiar to many researchers working in the relevant areas of general relativity.

Chapter four consists of the construction of, and investigation into, an ultra-local, Penrose-like limit. This chapter primarily consists of original work, some results of which have been published in \cite{Cropp2}.

The appendix provides the abstracts and publication details for both journal papers and the conference article that have resulted from the research presented in this thesis, all of which are coauthored with Prof. Matt Visser. 

\renewcommand{\cftchapfont}{%
  \fontsize{13}{-100}\bf\selectfont}
  
\renewcommand{\cftsecfont}{%
  \fontsize{12}{0}\rm\selectfont}  
  
\renewcommand{\cftsubsecfont}{%
  \fontsize{11}{0}\sl\selectfont}  
  
\titlespacing*{\chapter}{0mm}{-10pt}{30pt}

\tableofcontents

\titlespacing*{\chapter}{0mm}{-10pt}{40pt}

\mainmatter

 \chapter{Introduction to general relativity}

Here we will very briefly describe some of the important aspects of general relativity, focusing on techniques and results that will be important to the research presented in this thesis.

General relativity is a theory of gravity first formulated in around 1915, primarily by Einstein, with significant contributions from Hilbert, and which has been developed extensively in the intervening decades. In general relativity the force of gravity is due to spacetime curving in the presence of matter. The natural mathematical language for considering curved spaces is that of tensors and differential geometry. 

\section{Geometrical tensors}

\noindent
Spacetime is described by the four-dimensional metric tensor, $g_{ab}$, with the associated invariant interval
\begin{equation}
\d s^2 = g_{ab}\,\d x^a\, \d x^b.
\end{equation}
The metric tensor describes the infinitesimal notion of distance in various directions. The important geometrical quantities in general relativity are formed from the metric tensor and its derivatives.

In particular, when considering the acceleration due to gravity it is natural to consider tensors formed from up to second derivatives of the metric. There are a limited number of these.

Firstly, it is helpful to introduce the connexion/Christoffel symbol (\emph{not} a tensor), given by\index{Christoffel symbol}
\begin{equation}
\Gamma^{a}{}_{bc}=\frac{1}{2}\,g^{ad}\,\left( g_{db,c} + g_{dc,b} - g_{bc,d} \right).
\end{equation}
The Christoffel symbol acts to propagate a vector infinitesimally from one point to another, and can be defined in terms of the covariant derivative of a tensor
\begin{equation}
\nabla_b T^a = \partial_b\, T^a + \Gamma^a{}_{cb}\,T^c.
\end{equation}
Now the Riemann tensor can be formed\index{Riemann tensor} from the Christoffel symbols and their derivatives
\begin{equation}
R^a{}_{bcd}=\partial_{c}\,\Gamma^{a}{}_{bd} - \partial_{d}\,\Gamma^{a}{}_{bc} + \Gamma^{a}{}_{ec}\,\Gamma^{e}{}_{bd}-\Gamma^{a}{}_{ed}\,\Gamma^{e}{}_{bc}. 
\end{equation}
The Riemann tensor describes the curvature of the spacetime, and can be associated with tidal forces on nearby test particles. 

The Riemann tensor can be formed from two simpler tensors;
\begin{align}
R_{abcd}=C_{abcd} &+ \frac{1}{2}\left(g_{ac}\,R_{bd}+g_{bd}\,R_{ac}-g_{ad}\,R_{bc}-g_{bc}\,R_{ad}\right) \nonumber \\
 &- \frac{1}{6}R\left(g_{ac}g_{bd}-g_{ad}g_{bc}\right),
 \end{align}
where $C_{abcd}$ is known as the Weyl tensor, defined by the above equation, the significance of which is that it is the curvature which remains present in a vacuum, and 
\begin{equation}\index{Ricci tensor}
R_{ab}=R^c{}_{acb};
\end{equation}
\begin{equation}\index{Ricci tensor}
R=g^{ab}\,R_{ab},
\end{equation}
are the Ricci tensor and Ricci scalar, respectively. 

Finally, the Einstein tensor is constructed from the Ricci tensor and Ricci scalar,
\begin{equation}\index{Einstein tensor}
G_{ab}=R_{ab}-\frac{1}{2}R\,g_{ab}\,.
\end{equation}


\section{Geodesics}
\index{geodesic|(}

\noindent
Geodesics are the paths, often denoted by $\gamma$, followed by test particles, influenced only by the curve of spacetime. Here by a test particle we mean the mass of the particle does not affect the spacetime through which it travels. There are three sorts:

\begin{itemize}
\item spacelike, followed by hypothetical faster-than-light particles (tachyons) and therefore of lesser physical interest
\item lightlike, also called null, followed by particles travelling at the speed of light (luxons) 
\item timelike, followed by slower than light particles (tardyons/bradyons)
\end{itemize}
Understanding geodesics is of prime importance when trying to understand the physics of general relativity.

In familiar Newtonian physics, the path of a free particle is described by\index{Newtonian mechanics}
\begin{equation}
\frac{\d^2 x^a}{\d t^2}=0,
\end{equation}
indicating no forces act on the particle. Geodesic motion is similar; no forces act on the particle and it simply falls under the influence of the spacetime metric. In Newtonian mechanics, a free particle follows a straight line; likewise geodesics are the straightest possible paths, which are also the paths of shortest length in the appropriate sense.

 \index{geodesic equation}
To describe this straightest path we take a tangent vector at a point, $T^a$, and require that the tangent vector can be parallel transported so that the transported vector is proportional to the original, 
\begin{equation}\label{desic}
T^a\,\nabla_aT^b \propto T^b.
\end{equation}

Consider an arbitrary parameterisation of a curve $x^a(\lambda)$, which has the tangent $\displaystyle\frac{\d x^a}{\d \lambda}$. Insert these into equation (\ref{desic}) to give the geodesic equation\index{geodesic equation}
\begin{equation}\label{geo}
\frac{\d^2x^a}{\d \lambda^2}+\Gamma^a{}_{bc}\,\frac{\d x^b}{\d \lambda}\,\frac{\d x^c}{\d \lambda} = f(\lambda)\,\frac{\d x^a}{\d \lambda},
\end{equation}
with $f(\lambda)$ some arbitrary function. It is always possible to write this in an affine parameterisation by an appropriate choice of $\lambda$, which simplifies equation (\ref{geo}) to 
\begin{equation}
\frac{\d^2x^a}{\d \lambda^2}+\Gamma^a{}_{bc}\,\frac{\d x^b}{\d \lambda}\,\frac{\d x^c}{\d \lambda} = 0.
\end{equation}

Lightlike geodesics have, for any parameterisation\index{geodesic!null}
\begin{equation}
g_{ab}\, \frac{\d x^a}{\d \lambda}\, \frac{\d x^b}{\d \lambda} = 0.
\end{equation}
For timelike geodesics, an affine parameter can be chosen to satisfy\index{geodesic!timelike}
\begin{equation}
g_{ab}\, \frac{\d x^a}{\d \lambda}\, \frac{\d x^b}{\d \lambda} = -1.
\end{equation}
\index{geodesic|)}


\section{Coordinate independence}
\noindent
For every scenario there are many possible representations, many possible choices of coordinates. The choice of coordinates selected might have some advantage in ease of representing a particular feature of the spacetime. 

As a simple example of this, consider the Schwarzschild solution\index{Schwarzschild spacetime}, a non-rotating black hole. The standard choice of coordinates, often called Schwarz\-schild coordinates or curvature coordinates is
\begin{equation}
\d s^2 =-\left(1-\frac{2M}{r}\right)\,\d t^2 + \frac{1}{1-\frac{2M}{r}}\,\d r^2 +r^2\, \left(\d\theta^2+\sin^2(\theta)\,\d\phi^2\right).
\end{equation}
However, numerous other coordinate choices have been used over the years, such as for instance the Painleve-Gullstrand coordinates
\begin{equation}
\d s^2 = -\left(1-\frac{2M}{\bar{r}}\right)\,\d \bar{t}^2 +2\sqrt{\frac{2M}{\bar{r}}}\,\d \bar{r}\,\d \bar{t}+ \d \bar{r}^2 +\bar{r}^2 \left(\d\theta^2+\sin^2(\theta)\,\d\phi^2\right).
\end{equation}

Two metrics represent the same spacetime if there exists a transformation between the two. Often explicitly finding this transform is the best way to show two metrics are equivalent. 

There is a transformation of the time coordinate between these two forms. The major physical difference between these forms is that at the event horizon at $r=2M$, in Schwarzschild coordinates the $\d r$ component becomes infinite, resulting in what is known as a coordinate singularity, whereas the Painleve-Gullstrand coordinates can be extended across the horizon. However the Schwarzschild form has a simpler diagonal form, and the meaning of the time coordinate in the two forms is different. 

Since the best coordinates to use depends on what properties one wants to investigate, and in many cases different choices will be best for different properties, it good to understand (at least in principle) the basic construction of important features in different commonly used systems. 

\vspace{20pt}
\noindent
Often coordinates are chosen because they represent what is seen by an important class of observers. For instance, consider synchronous coordinates\index{synchronous coordinates}, which are of the form
\begin{equation}
\d s^2 = -  \d t^2 + g_{ij}(t,x) \; \d x^i \d x^j.
\end{equation}
This means that paths with fixed spatial coordinates, $x^i=x^i{}_0$, and varying $t$ are geodesics. The coordinate $t$ represents the proper time of these observers. As such synchronous coordinates describe what is seen by freely falling observers. Such coordinates often cannot be globally extended as there are coordinate singularities where two such paths cross. Thus it is not always possible to pick such a system globally, but in theory is always possible to do locally (though technically difficult in practice).

\section{Tetrads}\index{tetrads}
\noindent
In any coordinate system it is possible to pick what is known as a tetrad frame (also referred to as a vierbein, orthonormal basis, or non-coordinate basis).

To construct such a basis, at any point pick four linear independent vectors, $e^a{}_\alpha$ (here $\alpha$ is merely a way to number which vector we are talking about) that are orthonormal in the sense
\begin{equation}
g_{ab}\,e^a{}_\alpha\, e^b{}_\beta = \eta_{\alpha \beta},
\end{equation}
where
\begin{equation} 
\eta_{\alpha\beta}=\left[\begin{array}{cccc} -1&0&0&0\\ 0& 1 & 0&0\\ 0&0& 1&0\\0&0&0&1 \end{array} \right],
\end{equation}
is the metric of Minkowski space, appropriate for special relativity.

The matrix formed from these contravariant vectors has an inverse matrix which is formed from the covariant vectors $e^\alpha{}_a$. Given this set up, tetrads form an orthonormal basis to the tangent space at a point. This implies
\begin{equation}
g_{ab}=e^\alpha{}_a\, e^\beta{}_b\,\eta_{\alpha \beta}.
\end{equation}

Such objects are also useful in other dimensions, where in general these are known as vielbeins. Particularly useful are three dimensions (triads or dreibeins) and two dimensions (zweibeins). 

Tetrads are useful for comparing coordinate-based calculations to what would be observed in a given scenario, as each observer essentially carries a tetrad basis (their standard notion of space and time) with them. In particular to make useful comparisons to the matter/energy that would be observed, one should use a tetrad basis. Objects calculated in a tetrad frame (such as the stress-energy) will be denoted by hats over the indices. 


\section{Matter}\label{matter}
\noindent
In addition to tensorial descriptions of the spacetime and its properties we want tensorial descriptions of any matter that may be present in the spacetime. This is achieved with the stress-energy tensor:
\begin{align}\index{stress-energy tensor}
T_{ab}=\left[\begin{array}{cc}\rho	\quad &	F_j	\\
\\
F_i	\quad &	\pi_{ij}	\end{array} \right],
\end{align}
with $\rho$ the density, $F_i$ the flux of energy, and $ \pi_{ij}$ the stress tensor, which describes the pressures and stresses present.

Of course, the form of the stress-energy is coordinate dependent. For instance, an observer passing through a field of dust will notice a flux, whereas one moving with the dust will not. Several useful types of matter to examine are:

\begin{enumerate}

\item{Vacuum}

Vacuum has $T^{ab}=0$, independent of the coordinates used. \index{vacuum equations}

\item{Dust}\index{dust}

Dust is a from of matter that is without pressure, given by
\begin{equation}
T^{ab} = \rho V^a V^b,
\end{equation}
with $V^a$ timelike.

In a comoving orthonormal frame it is simply given by 
\begin{equation}
T^{\hat a \hat b} =\left[\begin{array}{cccc} \rho&0&0&0\\ 0& 0 & 0&0\\ 0&0& 0 &0\\0&0&0&0 \end{array} \right].
\end{equation}

\item{Null dust}\index{null dust}

Null dust is the lightlike version of pressure-less matter, suitable for describing photons and hypothetical speed-of-light particles. The stress-energy tensor can be decomposed into the product of a null vector with itself, similarly to the dust case,
\begin{equation}
T^{ab}=k^ak^b.
\end{equation}
If we employ null coordinates
\begin{equation}
u=z-t; \qquad v=z+t,
\end{equation}
there is a frame where the only component is $T_{uu}$.

\item{Perfect fluid}\index{perfect fluid}

A perfect fluid has density and pressure, but no viscosity, and hence cannot support a shear, and additionally all pressures are the same. A perfect fluid is described by
\begin{equation}\label{pfluid}
T^{ab} = (\rho+p) V^a V^b + p g^{ab},
\end{equation}
so in a comoving orthonormal frame it takes the form
\begin{equation}
T^{\hat a\hat b} =\left[\begin{array}{cccc} \rho&0&0&0\\ 0& p & 0&0\\ 0&0& p&0\\0&0&0&p \end{array} \right].
\end{equation}

\item{Scalar field}\index{scalar field}\label{scalar}

The stress-energy of a scalar field can be expressed as
\begin{equation}
T_{ab} = \phi_{,a}\, \phi_{,b} - \frac{1}{2}\, g_{ab}\, \left\{ (\nabla\,\phi)^2 + V(\phi) \right\}.
\end{equation}

Examples of (hypothetical) fundamental scalar fields are the Higgs field and dilatons.

\item Electromagnetic field\index{electromagnetic field}

The electromagnetic stress-energy tensor is given by 
\begin{align}
\nonumber \\
T_{ab}=\left[\begin{array}{cc}\frac{1}{2}\,(E^2+B^2)	\quad &	\vec E \times \vec B	\\
\nonumber \\
\vec E \times \vec B	\quad &	E_i\,E_j+B_i\,B_j-\frac{1}{2}\,(E^2+B^2)\delta_{ij}	\end{array} \right],
\nonumber \\
\end{align}
containing the familiar Maxwell stress tensor, and the Poynting vector as flux.

It can sometimes be convenient to express fields using the vector potential, $A^a$,\index{vector potential}
\begin{equation}\label{vecpot}
A^a(t,x) = \left( \phi(t, x); \; \vec A(t,x) \right),
\end{equation}
which describes the electric and magnetic fields through
\begin{equation}
\vec B= {}^{}\nabla\times \vec A; \qquad \vec E = -\nabla \phi - \frac{\partial \vec A}{\partial t}.
\end{equation}
We can express the electromagnetic field tensor as\index{electromagnetic field tensor}
\begin{equation}
F_{ab}=A_{b;a}-A_{a;b},
\end{equation}
or explicitly,
\begin{align}
\nonumber \\
F_{ab}=\left[\begin{array}{cccc} 0&E_x&E_y&E_z\\ -E_x& 0 & B_z&-B_y\\ -E_y&-B_z& 0 &B_x\\-E_z&B_y&-B_x&0 \end{array} \right].
\nonumber \\
\end{align}
In terms of the electromagnetic field tensor, the stress-energy is given by
\begin{equation}
T_{ab}=F_{ac}\,g^{cd}F_{db}-\frac{1}{4}\,g_{ab}\left(F_{ef}\,F^{ef}\right).
\end{equation}
Note the trace of the stress-energy vanishes, $\displaystyle g^{ab}\,T_{ab} = 0$.

\end{enumerate}


\section{Einstein equations}\index{Einstein equations}
\noindent
This brings us to the central equations of general relativity, the Einstein equations:
\begin{equation}
G_{ab}=8\pi G_N \,T_{ab}.
\end{equation}
These equations link the presence of matter to the curvature present in the spacetime, and determine how that curvature will evolve. 

Although simple in appearance these are ten second-order non-linear partial differential equations in the general case, which have only been solved exactly in situations of high symmetry \cite{Podolskyexact, kramerexact}.
\index{conservation law}\index{Bianchi identity}

However, these ten equations are not independent, and are constrained by the Bianchi identity, which, in the context of general relativity (rather than abstract differential geometry in general), is
\begin{equation}
\nabla_a G^{ab}=0.
\end{equation}
This constraint means that only six of the ten of the Einstein equations are independent, and we thus have coordinate freedom to reduce the equations down to six. 

The Bianchi identity can also be regarded as a conservation relation on the stress-energy tensor,
\begin{equation}
\nabla_a T^{ab}=0,
\end{equation}
as a consequence of the Einstein equations.

\vspace{20pt}
\noindent
For vacuum the Einstein equations reduce to a statement about the Ricci tensor as\index{vacuum equations}
\begin{equation}
R_{ab}-\frac{1}{2}\,R\,g_{ab}=0 \quad \Rightarrow \quad R_{ab}=0,
\end{equation}
which gives the vacuum equations. 


\section{Weak-field gravity}\index{weak-field}\index{linearised gravity}
\noindent
Although solving the general equations in situations without significant symmetry is difficult, it is possible to explore the weak-field, linearised results more generally. This was one aspect that was investigated soon after the development of general relativity, primarily to predict differences from Newtonian gravity. \index{Newtonian mechanics}

When a gravitational field is weak it is expected that the spacetime metric should be close to the description of special relativity;
\begin{equation}
\d s^2 =\eta_{ab}\,\d x^a\, \d x^b.
\end{equation}
So the linearised theory of of gravity is obtained by using the metric:
\begin{equation}
g_{ab}=\eta_{ab}+h_{ab}; \qquad h_{ab} \ll 1,
\end{equation}
and neglecting all second order or higher terms when calculating relevant physical variables, for instance
\begin{equation}
\Gamma^a{}_{bc}=\frac{1}{2}\,\eta^{ad}\,\left(h_{db,c}+h_{dc,b}-h_{bc,d}\right)+O(h^2),
\end{equation}
\begin{equation}\label{ricciwave}
R_{ab}=\frac{1}{2} \left(h^c{}_{a,bc}+h^c{}_{b,ac}-\nabla^2\, h_{ab}-h_{,ab}\right)+O(h^2).
\end{equation}
In addition we have gauge freedom to set
\begin{equation}
\left(h^c{}_b-\frac{1}{2}\,h\,\delta^c{}_b\right)_{,c}=0.
\end{equation}
This is known variously as the Einstein/Hilbert/de Donder/Fock gauge, and simplifies equation (\ref{ricciwave}) down to\index{gauge condition}
\begin{equation}\label{ricciwavesol}
R_{ab}=-\frac{1}{2}\, \nabla^2\, h_{ab}+O(h^2),
\end{equation}
and consequently,
\begin{equation}
G_{ab}=-\frac{1}{2}\,\nabla^2 \left\{h_{ab}-\frac{1}{2}\,h\,\eta_{ab} \right\}+O(h^2).
\end{equation}
So we have reduced the Einstein equations to simpler, linear differential equations. 

Note the strongest gravitational fields in our Solar System are well within the regime where these approximations are a very good description of reality. This means that weak-field gravity has been very important for measurable predictions and applications. These include time dilation due to gravity wells, which is important in GPS applications and the precession of perihelion of Mercury that had been noted prior to the development of general relativity. Another important prediction is that gravity propagates through spacetime as a wave, direct evidence for which is still being sought. We will examine these weak-field waves in some detail. 

\subsection{Weak-field gravitational waves}\index{gravitational waves!weak-field}
\noindent
From equation (\ref{ricciwavesol}) it is clear that the vacuum solutions for weak-field gravity are\index{vacuum equations}\index{wave equation}
\begin{equation}
\nabla^2\, h_{ab}=0.
\end{equation}
This means that the metric obeys the wave equation in a vacuum, and the effects of changes in gravitational fields propagate via gravitational waves. 
A way to note the physical effect of these waves is to look at the (weak-field) Riemann tensor, \index{Riemann tensor}
\begin{equation}
R_{abcd}=\frac{1}{2}\, \left\{ h_{ad,bc} + h_{bc,ad} - h_{ac,bd} - h_{bd,ac} \right\},
\end{equation}
and so
\begin{equation}
\nabla^2\,R_{abcd}=0. 
\end{equation}
We can investigate the properties of gravitational waves through their effects on test particles due to tidal forces apparent by the non-zero Riemann tensor. 

\vspace{20pt}

\noindent
Consider a plane wave travelling in the $z$ direction, which we could write as 
\begin{equation}
\d s^2 = -\d t^2 + \d z^2 + \left(\delta_{AB}+h_{AB}(t, z)\right)\d x^A\, \d x^B,
\end{equation}
or alternatively, 
\begin{equation}\label{nuwave}
\d s^2 = \d u\,\d v + \left(\delta_{AB}+h_{AB}(u)\,\right)\d x^A\, \d x^B,
\end{equation}
where we are employing null coordinates:
\begin{equation}
u=z-t; \qquad v=z+t.
\end{equation}
Note there are several natural definitions for null coordinates that are commonly seen, such as
\begin{equation}
u=\frac{1}{\sqrt{2}}(z-t); \qquad v=\frac{1}{\sqrt{2}}(z+t),
\end{equation}
\begin{equation}
u=t-z; \qquad v=t+z.
\end{equation}
These minor differences lead to frequently seen differences in factors of $\pm 1$ or $\pm 2$.

For the wave given in equation (\ref{nuwave}) the components of the Riemann tensor are $R_{uyuy}$, $R_{uxux}$, and $R_{uxuy}$, so there is no motion of test particles in the $z$ direction. Therefore we will consider a circle of test particles perpendicular to the $z$-axis. It is found that the circle is distorted whilst maintaining the same area. 

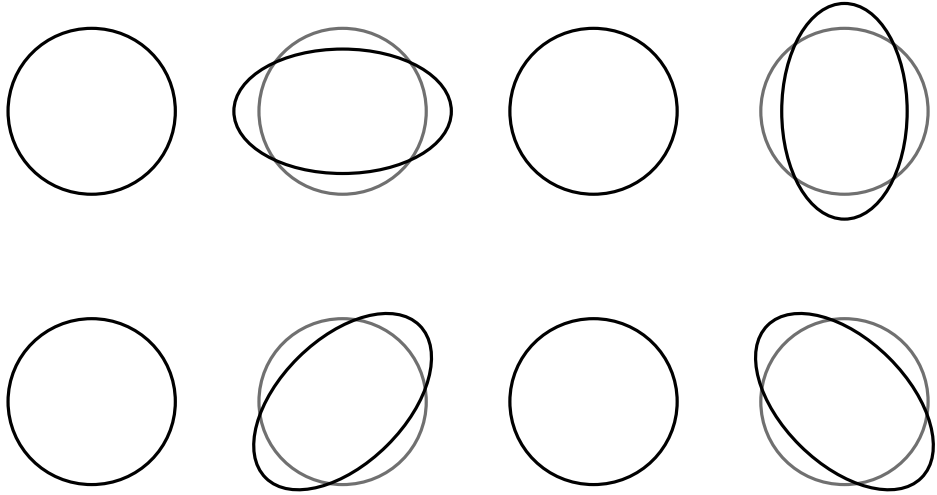
\begin{figure}
\begin{center}
\begin{tikzpicture}[scale=1.1]
\draw (0,0) [very thick, black]ellipse (1cm and 1cm);
\draw [very thick, gray!40!black, opacity=0.7, xshift=3cm]ellipse (1cm and 1cm);
\draw [very thick, black, xshift=3cm] ellipse(1.3cm and 0.75cm);
\draw [xshift=6cm, very thick]ellipse(1cm and 1cm);
\draw [very thick, gray!40!black, opacity=0.7, xshift=9cm]ellipse(1cm and 1cm);
\draw [xshift=9cm, very thick]ellipse(0.75cm and 1.3cm);
\draw[yshift=-3.5cm, very thick]ellipse (1cm and 1cm);
\draw [very thick, gray!40!black, opacity=0.7, xshift=3cm,yshift=-3.5cm]ellipse (1cm and 1cm);
\draw[yshift=-3.5cm, xshift=3cm, rotate=45, very thick]ellipse(1.3cm and 0.75cm);
\draw[xshift=6cm, yshift=-3.5cm, very thick]ellipse(1cm and 1cm);
\draw [very thick, gray!40!black, opacity=0.7, xshift=9cm, yshift=-3.5cm]ellipse(1cm and 1cm);
\draw[xshift=9cm, yshift=-3.5cm,rotate=45, very thick]ellipse(0.75cm and 1.3cm);
\end{tikzpicture}
\end{center}
\caption{The effect of the $+$ (upper) and $\times$ (lower) polarized gravitational waves, on a circle of particles with the plane of the page perpendicular to the direction of propagation}
\label{fig:waves}
\vspace{12pt}
\end{figure}

To concisely describe the way that such a circle of particles is distorted we introduce polarisation modes. For gravitational waves the two modes are 45$^\circ$ out of phase. They are referred to as the $+$ and $\times$ polarisations. The effect of these are shown in figure (\ref{fig:waves}). Pure $+$ polarisation can be written as \index{gravitational waves!linear polarisation}
\begin{equation}
\d s^2 = \d u\, \d v + \d x^2 + \d y^2 +\frac{1}{2}\,f(u)\,\left(\d x^2 -\d y^2 \right),
\end{equation}
and the $\times$ polarisation as 
\begin{equation}
\d s^2 = \d u\, \d v + \d x^2 + \d y^2 +g(u)\,\d x \d y.
\end{equation}

As superposition is possible due to the linearity of the wave equation, these two modes can simply be combined to create any weak-field gravitational wave, with some appropriate choice of $f(u)$ and $g(u)$:
\begin{equation}\label{weak:general}
\d s^2 = \d u\d v + \d x^2 + \d y^2 + \frac{f(u)}{2}\,\left(\d x^2 - \d y^2\right) + g(u)\,\d x \d y,
\end{equation}
where in all cases $f(u) \ll 1, g(u) \ll 1$.

For instance, a circularly polarized wave can be constructed by
\begin{equation}\label{weak:circular}
\d s^2 = \d u \d v + \d x^2 + \d y^2 +S\left(\cos(ku)\,\frac{\left(\d x^2 -\d y^2 \right)}{2}+ \sin(ku)\,\d x \d y\right),
\end{equation}
with $S \ll 1$ some constant.

\section{The Raychaudhuri equation}
\label{raychaud}
\index{Raychaudhuri equation}
\noindent
In general relativity the Raychaudhuri equation \cite{HE, Wald} is used to describe the motion of nearby bits of matter and whether they converge or not. There are two versions, for lightlike and timelike particles. For the timelike case one starts with a collection of nearby timelike vectors, $u^a$ (a congruence). An example of this would be the 4-velocity of a fluid. Define the projection operator

\begin{equation}\label{intro:a}
h_{ab}=g_{ab}+u_a\,u_b.
\end{equation}

Define the expansion tensor and scalar, \index{expansion}
\begin{equation}
\theta_{ab}=h_{ac}\,\nabla^{(c}u^{d)}\,h_{db},
\end{equation}
\begin{equation}
\theta=g^{ab}\,\theta_{ab}= \nabla_a\, u^a.
\end{equation}
The expansion tensor measures the rate at which the volume of a small ball of fluid changes. 
The shear tensor and scalar are:\index{shear}
\begin{equation}
\sigma_{ab}=\theta_{ab}-\frac{1}{3}\,h_{ab}\,\theta,
\end{equation}
\begin{equation}
\sigma^2=\frac{1}{2}\,\sigma_{ab}\,\sigma^{ab} \geq 0.
\end{equation}
These measure the tendency for a small ball of fluid to be distorted into an ellipsoid shape.
Now consider the vorticity tensor and scalar, \index{vorticity}
\begin{equation}
\omega_{ab}=h_{ac}\,\nabla^{[c}u^{d]}\,h_{db},
\end{equation}
\begin{equation}
\omega^2=\frac{1}{2}\,\omega_{ab}\,\omega^{ab} \geq 0.
\end{equation}
These measure how the nearby vectors twist around each other.

Note there exist some minor differences in commonly used notation. We are following the conventions of Hawking and Ellis \cite{HE}.

The expansion, shear and vorticity tensors can also be viewed as a decomposition of $ \nabla_b\, u_a$ into a symmetric traceless tensor (shear), symmetric tensor with trace (expansion), and an antisymmetric tensor (vorticity). That is,
\begin{equation}\label{ray1}
\nabla_b\, u_a = \omega_{ab} +\sigma_{ab}+\frac{1}{3}\,\theta\, h_{ab} - \frac{\d u_a}{\d s}\,u_b.
\end{equation}

Raychaudhuri's equation is
\begin{equation}
\frac{\d \theta}{\d s}= -R_{ab}\,u^a\,u^b+2\omega^2 -2\sigma^2 -\frac{1}{3}\,\theta^2+\nabla_{a}\,\left(\frac{\d u^a}{\d s}\right).
\end{equation}

The lightlike case is similar, except that the definition of the projection operator $h_{ab}$ as in equation (\ref{intro:a}) does not work in quite the same way, due to technical issues with constructing orthogonal vectors to a null ray. This is because the null ray has zero length, hence is orthogonal to itself.

Due to the physically useful picture these terms provide it is often of interest to describe the form of the vorticity, shear and expansion tensors in particular spacetimes, especially if the Raychaudhuri equation simplifies in some way. The major use for this equation has been in the development of the singularity theorems \cite{HE}. It has also been of use in such diverse contexts as an astrophysical understanding of when cracks form in dense objects (see section IV of \cite{kar} and the references therein), and of importance in understanding the gravitational analogues of the laws of thermodynamics. The lightlike case is particularly useful in examining gravitational lensing. Various extensions have been made to this equation in cases such as speculative theories of gravitation involving torsion. See \cite{Ellisray, kar} for reviews.

\section{Splitting of spacetime into space and time}\label{ADM}
\noindent
The Arnowitt, Deser and Misner (ADM) decomposition \cite{ADM, MTW} is a way of splitting a general metric into propagating slices or ``leaves''. This splitting is referred to as a foliation. The spacetime metric is decomposed into the lapse function, $N$, shift vector, $N^i$, and spatial metric $g_{ij}$;
\begin{align}
g_{ab}=\left[\begin{array}{c|c}
\mbox {$N_kN^k-N^2$} \quad & \quad \mbox{$N_k$} \quad \\ 
\hline
\mbox{$N_k$} \quad & \quad \mbox{$g_{ij}$} \quad \\
\end{array} \right].
\end{align}

Then the inverse metric is 
\small
\begin{align}
\displaystyle g^{ab}=\left[\begin{array}{c|c}
\\
\quad \displaystyle-\frac{1}{N^2} \quad &\displaystyle \frac{N^k}{N^2} \\ 
\\
\hline
\\
\quad \displaystyle \frac{N^k}{N^2} \quad &\displaystyle g^{ij} - N^i\,N^j/N^2 \\
\\
\end{array} \right].
\end{align}
\normalsize

This decomposition has been used in various attempts to construct a quantum theory of gravity. The construction is also useful for understanding one of the common concepts of mass in general relativity, the ADM mass. 

\subsection{Extrinsic and intrinsic curvature}\index{extrinsic curvature}\index{intrinsic curvature}
\noindent
There are two sorts of curvature that may be of mathematical and physical importance. General relativity primarily concerns itself with the \emph{intrinsic} curvature (of four-dimensional spacetime), which is the curvature which can be measured by observers stuck inside the space itself, and is measured by the Riemann tensor. Another important way of looking at curvature is the \emph{extrinsic} curvature, which is the curvature measured with respect to an embedding in a higher-dimensional space. The most notable simple difference between these two concepts is for an arbitrary 1D space, which has zero intrinsic curvature (the Riemann tensor is always zero in 1D) but which can be defined as curving by placing it in a 2D space and ascribing a radius of curvature to the various points on the curve. 

Let us make these two concepts precise: take a foliation of spacetime, and take a normal vector, $\mathcal{N}$, along some path, $s$, in a given slice. This normal projects a slice infinitesimally into the future slice. The extrinsic curvature, $K$, measures how the normal changes direction from one point on a slice to another. 
\begin{equation}
K=\frac{\d \mathcal{N}}{\d s}.
\end{equation}

Although the intrinsic curvature is of most interest in general relativity, we saw above that it is possible to split 4D spacetime into 3D spatial slices propagating in a fourth, timelike, dimension. This means we can discuss the extrinsic curvature of the spatial slices with respect to the timelike coordinate. In fact, it is always possible to express intrinsic four dimensional curvature in terms of the extrinsic curvature and the intrinsic curvature of the spatial slices \cite{MTW}. Depending on the particular spacetime considered this may be a very useful property. 

This can be particularly useful when one is working in the ADM viewpoint, as we may calculate the extrinsic curvature as
\begin{equation}
K_{ij}=\frac{1}{2N}\,\left(N_{i;j}+N_{j;i}-\frac{\partial g_{ij}}{\partial t}\right),
\end{equation}
where the covariant differentiation is with respect to the three dimensional metric, $g_{ij}$. 

\subsection{Lagrangian formulation}\index{Lagrangian}
\noindent
The Lagrangian for (vacuum) general relativity is 
\begin{equation}
\mathcal{L} = \sqrt{-{}^{(4)}g}\,{}^{(4)}R.
\end{equation}
In the ADM viewpoint there is a simple expression for the Lagrangian
\begin{equation}
\mathcal{L} =  \;{}^{(3)}g^{1/2}\,N\,\left(K_{ij}\,K^{ij} - K^2+{{}^{(3)}}R\right).
\end{equation}
This means that we have the simple interpretation that 
\begin{equation}
K_{ij}\,K^{ij} - K^2
\end{equation}
can be regarded as kinetic energy, and ${}^{(3)}R$ as the potential energy.

\section{Discussion}
\noindent
In this chapter we have outlined the basics and some key important results from the theory of general relativity. We have been highly selective, focusing only on items that are relevant to the rest of this thesis. 

The Einstein equations link the evolution of spacetime with the matter content, and we can describe these spacetimes with a metric tensor. Furthermore even in the case of weak-field gravity, general relativity makes some wildly different predictions form Newtonian gravity, such as the existence of gravitational waves. 

In addition we saw a useful way to split spacetime and discovered some nice ways to examine the behaviour of matter, through the geodesic equations and the Raychaudhuri equations. 
\chapter{Polarisations of the Rosen form}\label{rosenchap}

In the introduction we looked at weak-field gravitational waves. Now we will consider their strong-field analogues, in the two common forms they are presented in, the Rosen and Brinkmann forms. We find in one of these forms it is a simple matter to write down any arbitrary polarisation, while in the other form it is not clear how to construct such polarisations.

To address this issue we develop an algorithm to write down arbitrary polarisations, filling this gap in our understanding of strong-field gravitational waves. The research in this chapter has been presented in \cite{Cropp1, Cropp3}.

\section{Strong-field gravitational waves}\index{gravitational waves!strong-field}

We have a complete description of weak-field gravitational waves, which will include every gravitational wave we can potentially directly detect on Earth. However strong-field, non-linear, gravity is important in many astronomical events, which will produce gravitational radiation, and thus it is important to understand the strong-field gravitational waves. In general this is a very complex problem, so one aspect that is commonly investigated is the case with planar symmetry. While this is an unrealistic assumption, it can be hoped that some general features of strong-field gravitational waves may be understood by understanding this simplest case, and the planar case already demonstrates some interesting features and raises some important questions. 

Note that strong-field gravitational waves do \emph{not} obey the wave equation: this is not a problem, and is not unique to gravitation, instead being a reflection on the assumptions which are necessary to obtain the wave equation. This means strong-field waves do not obey superposition, as the Einstein equations are not linear, and the investigation of colliding gravitational waves has been of considerable interest within the relativity community \cite{chandra, Podolskyexact}.

Progress was made in finding the strong-field analogues of weak-field gravitational waves, in situations of high symmetry, by Einstein and Rosen \cite{Rosen} and independently by Brinkmann \cite{Brinkmann}, in two different forms. We shall examine both these forms of strong-field gravitational plane waves, starting by introducing a specific ansatz based on our knowledge of weak-field waves. Finally we shall show that the two forms are related by a coordinate transform, and hence describe the same phenomenon. The respective advantages and disadvantages of the two forms will be mentioned. 


\subsection{Rosen form}\label{pp:rosen}\index{gravitational waves!Rosen form}

We have found we can write weak-field waves as
\begin{equation}
\d s^2=\d u\,\d v+(\delta_{AB}+h_{AB}(u))\d x^A\,\d x^B.
\end{equation}
One ansatz for a strong field wave is simply to remove the restriction that $h_{AB}(u)$ is small, so
\begin{equation}
\d s^2=\d u\,\d v+g_{AB}(u)\d x^A\,\d x^B.
\end{equation}
It is then a standard result \cite{kramerexact} that the only non-zero component of the Ricci tensor is:
\begin{equation}
R_{uu} = - \left\{ \frac{1}{2} \; g^{AB} \; g_{AB}'' 
- \frac{1}{4} \; g^{AB} \; g_{BC}' \; g^{CD} \; g_{DA}' \right\} =G_{uu},
\end{equation}
as $R=0$.

Vacuum solutions for this metric are strong field gravitational waves in the Rosen form. Note that for weak-field waves $h_{AB}(u)$ was a completely arbitrary symmetric matrix function of $u$, while in the strong-field case we require $R_{uu}=0$, that is, 
\begin{equation}
 \frac{1}{2} \; g^{AB} \; g_{AB}'' 
- \frac{1}{4} \; g^{AB} \; g_{BC}' \; g^{CD} \; g_{DA}' =0,
\end{equation}
so we have a restriction on $g_{AB}$.\index{vacuum equations}

It is easily found, up to the usual symmetries, that for a wave travelling in the $z$-direction, the only non-zero components of the Riemann tensor are $R_{uxux}$, $R_{uxuy}$ and $R_{uyuy}$, given by \cite{Cropp1}\index{Riemann tensor}
\begin{equation}\label{pp:e}
R_{uAuB} = - \left\{ \frac{1}{2} \; g_{AB}'' 
- \frac{1}{4} \; \; g_{AC}' \; g^{CD} \; g_{DB}' \right\}.
\end{equation}

Considering some of the properties of this form we note that it is explicitly translation invariant in the directions perpendicular to motion, as expected from the symmetries for a plane wave. However this form is unfortunately prone to coordinate singularities which indicate the formation of caustics of geodesics.

\subsection{Brinkmann form}\label{pp:brinkmann}\index{gravitational waves!Brinkmann form}

Consider the metric
\begin{equation}
 \d s^2 = \d u \,\d v + \left\{ H_{AB}(u)\, x^A\, x^B\right\} \, \d u^2 + \d x^2 + \d y^2.
\end{equation}
This is the form for a plane wave in the Brinkmann form \cite{Brinkmann}. The only nonzero component of the Ricci and Einstein tensors is given by
\begin{equation}\label{brink}
R_{uu}=-\frac{1}{2}\,\left(H_{xx}(u)+H_{yy}(u)\right)=G_{uu},
\end{equation}
again, with $R=0$.

Furthermore (up to the usual symmetries) the non-zero components of the Riemann tensor are\index{Riemann tensor}
\begin{equation}\label{pp:f}
R_{uxux} =-\frac{1}{2}\,H_{xx}(u);
\end{equation}
\begin{equation}\label{pp:g}
R_{uyuy} =-\frac{1}{2}\,H_{yy}(u);
\end{equation}
\begin{equation}\label{pp:h}
R_{uxuy} =-\frac{1}{2}\,H_{xy}(u).
\end{equation}
Note that here, and with equation (\ref{brink}), the subscripts label components, and do not indicate derivatives. 

We see that this is the same pattern of non-zero components as for the Rosen form. This means that \emph{some} of our intuition, developed with weak-field Rosen form, can be carried over to strong-field Brinkmann form \cite{Podolskyexact}. 

Note this form is explicitly \emph{not} translation invariant in the $x$ and $y$ directions, unlike the Rosen form. However, this form has a distinct advantage in that it is not plagued by coordinate singularities in the same way that the Rosen form is. Thus we see that one form of strong-field gravitational waves better represents the symmetry of the system at the cost of the coordinates being inextendable beyond a certain point. 

Gravitational plane waves are then given by solving the algebraic vacuum equations
\begin{equation}
 H_{xx}(u)=-H_{yy}(u),
\end{equation}
or more explicitly, 
\begin{equation}
H_{AB} = \left[\begin{array}{cc}
H_+(u) & H_\times(u) \\ 
H_\times(u) & -H_+(u) \\
\end{array} \right],
\end{equation}
with $H_+(u)$ and $H_\times(u)$ completely arbitrary (the reason for this choice of labelling for these functions will become clear shortly).


\subsection{Transformation between the two forms}\index{gravitational waves!Rosen form}\index{gravitational waves!Brinkmann form}\index{coordinate transform}

To show these two forms really represent the same phenomenon, we will now demonstrate the coordinate transformation that takes the Brinkmann form into the Rosen form (see \cite{blaulecture2, kramerexact}). 

Starting with the Rosen form
\begin{equation}
\d s^2=\d U\,\d V+g_{AB}(U)\,\d y^A\,\d y^B,
\end{equation}
we will transform to the Brinkmann form:
\begin{equation}
\d s^2 = \d u \, \d v + H_{\alpha \beta}(u)\, x^\alpha x^\beta\, \d u^2+ \d \vec x^2.
\end{equation}
Begin by transforming the transverse coordinates as 
\begin{equation}
\label{3:a}
x^\alpha=E^\alpha{}_B\,y^B,
\end{equation}
where we use $x^\alpha$ for transverse coordinates in Brinkmann form, and $y^A$ for those in Rosen form, and $E^\alpha{}_B$ obeys
\begin{equation}
g_{AB}(U)=E^\alpha{}_A\,E^\beta{}_B\,\delta_{\alpha \beta}.
\end{equation}
So it is a zweibein. Now
\begin{equation}\label{pp:a}
\d y^A=E^A{}_\alpha\, \d x^\alpha+\dot E^A{}_\beta\, x^\beta \d U,
\end{equation}
and
\begin{equation}
g_{AB}\,\d y^A\,\d y^B=E^\alpha{}_A\,E^\beta{}_B\,\delta_{\alpha \beta}(E^A{}_\gamma \,\d x^\gamma+\dot E^A{}_\gamma\, x^\gamma \d U)(E^B{}_\delta\, \d x^\delta+\dot E^B{}_\delta\, x^\delta \d U).
\end{equation}
Now simplifying
\begin{equation}
g_{AB}\,\d y^A\,\d y^B=\delta_{\alpha \beta}(\d x^\alpha+\dot E^\alpha{}_A\, E^A{}_\gamma\, x^\gamma\, \d U)(\d x^\beta+\dot E^\beta{}_B\, E^B{}_\delta\, x^\delta\, \d U),
\end{equation}
and expanding out
\begin{align}
g_{AB}\,\d y^A\,\d y^B &=\, \delta_{\alpha \beta}\left(\d x^\alpha \d x^\beta + \dot E^\beta{}_B\, E^B{}_\delta\, x^\delta\, \d U\d x^\alpha \right. \nonumber \\
&+ \left. \dot E^\alpha{}_A\, E^A{}_\gamma\, x^\gamma\, \d U\d x^\beta +\dot E^\alpha{}_A\, E^A{}_\gamma\, x^\gamma\,\dot E^\beta{}_B\, E^B{}_\delta\, x^\delta\, \d U^2 \right).
\end{align}
This gives us a $dU^2$ term, and the desired form for the $dx^\alpha\,dx^\beta$ components, but with undesired $\d x^\alpha\d U$ terms. To complete the transformation we transform $v$:
\begin{equation}
V=v+\dot E_{\alpha A}\,E^A{}_\beta\, x^\alpha\, x^\beta=\frac{1}{2}\dot g_{AB}\,y^A\,y^B;
\end{equation}
\begin{equation}
\d V=\d v-\frac{1}{2}\,\ddot g_{AB}\,y^A\,y^B-\frac{1}{2}\,\dot g_{AB}\,\d y^A\, y^B-\frac{1}{2}\,\dot g_{AB}\,y^A\, \d y^B.
\end{equation}
Now using equation (\ref{pp:a}) to write $y^A$, $dy^A$ in terms of $x^\alpha$ and $E^\alpha {}_{A}$ we see the $\d x\d u$ terms cancel as well as producing new $du^2$ terms, giving us the desired Brinkmann form;
\begin{equation}
\d s^2= \d U \, \d V +\d \vec x^2 + \ddot E_{\alpha i}\, E^i{}_\beta\,x^\alpha\, x^\beta\, \d U^2. 
\end{equation}

\vspace{20pt}

\noindent
So the desired transformation from one coordinate system to another is:
\begin{eqnarray}
U &=& u\,;
\nonumber\\
y^A &=& E^A{}_\alpha\, x^\alpha;
\nonumber\\
V &=& v+\dot E_{\beta A}\,E^A{}_\alpha\, x^\beta\, x^\alpha.
\end{eqnarray}
It is thus seen that these two forms are related by a coordinate transform, hence represent the same phenomenon. Note that this transformation is singular at any point the Rosen form has a coordinate singularity.


\section{The {\it pp} waves}
There is a generalisation of the concept of gravitational plane waves; the {\it pp} waves.\index{pp@\textit{pp} spacetime}
Consider the spacetime geometry \cite{Penrose1, Penrose2, steele, kramerexact} 
\begin{equation}
\d s^2 = \d u \, \d v + H(u,x,y) \, \d u^2+ \d x^2 + \d y^2. 
\end{equation}
This is the Brinkmann form for a general {\it pp} spacetime.

It is then a standard result that the only nonzero component of the Ricci tensor is
\begin{equation}
R_{uu}= -\frac{1}{2} \left\{ \partial_x^2\, H(u,x,y) + \partial_y^2\,H(u,x,y) \right\} = G_{uu}.
\end{equation}
Where again $R=0$. Furthermore (up to the usual index symmetries) the only non-zero components of the Riemann tensor are of the form $R_{uAuB}$. Specifically:\index{Riemann tensor}
\begin{equation}
R_{uxux} = -\frac{1}{2}\, \partial_x^2\, H(u,x,y); 
\end{equation}
\begin{equation}
R_{uxuy} = -\frac{1}{2}\, \partial_x\, \partial_y\, H(u,x,y) ;
\end{equation}
\begin{equation}
R_{uyuy} = -\frac{1}{2}\, \partial_y^2\, H(u,x,y) . 
\end{equation}

This metric describes {\it pp} wave (plane-fronted waves with parallel propagation) spacetimes, of which plane gravitational waves are a subset \cite{Ehlers}. This spacetime describes gravitational waves where the source is either electromagnetic waves, waves of any hypothetical zero-mass particle, or null dust. 

One notable aspect of {\it pp} waves is that \cite{blaulecture2} all the scalar invariants vanish (such a spacetime is referred to as a vanishing scalar invariant (VSI) spacetime). This means a {\it pp} spacetime cannot have a scalar curvature singularity (where the curvature invariants become infinite), though it may have other sorts of singularities.

In addition to plane waves, some important examples of {\it pp} wave spacetimes are the Aichelburg-Sexl ultraboost \cite{Aichelburg} and the Bonnor beam \cite{Bonnor}.
The {\it pp} spacetimes are studied not just in the context of general relativity but also as purely mathematical constructs in studies of Lorentzian geometry and are of interest in string theory due to them often being exactly solvable backgrounds. Further they have generated interest due to the causal properties of these spacetimes \cite{Hubeny2, Penrose1}.

As we have seen, purely gravitational plane waves, as opposed to the more general {\it pp}-waves \cite{Penrose1, Penrose2, steele}, can be characterised by the fact that $H(u,x,y)$ is a quadratic function of the transverse coordinates \cite{Brinkmann, D'Inverno, steele}.


\section{Polarisations of the two forms}\label{pp:arbit}

We saw in section (\ref{pp:brinkmann}) that the general form for a plane gravitational wave in the Brinkmann form is 
\begin{equation}
\label{pp:b}
 \d s^2 = \, \d u \, \d v + \left\{ [ x^2-y^2] \, H_+(u) + 2xy \, H_\times(u) \right\} \, \d u^2+ \d x^2 + \d y^2. 
\end{equation}
We see $H_{+}(u)$ and $H_{\times}(u)$ are clearly the $+$ and $\times$ components of the polarisation for the wave. Thus in this form for the metric the two polarisation modes are explicitly seen to decouple and to superimpose linearly --- quite similarly to the situation in Maxwell electromagnetism --- and by choosing $H_+(u)$ and $H_\times(u)$ appropriately we can construct not just $+$ and $\times$ polarized waves, but also circular polarisation, elliptic polarisation, and even more general polarisation states. As a simple example, circular polarisation would be given by
\begin{equation}
\d s^2 = \, \d u \, \d v + H(u)\,\left\{ [ x^2-y^2] \, \cos(\Omega u) + 2xy \, \sin(\Omega u) \right\} \, \d u^2+ \d x^2 + \d y^2. 
\end{equation}

Equivalently, changing to polar coordinates, in the $x$--$y$ plane
\begin{equation}
\label{pp:c}
 \d s^2 = \d u \, \d v + r^2\left\{ \cos(2\phi) \, H_+(u) + \sin(2\phi) \, H_\times(u) \right\} \, \d u^2+ \d r^2 + r^2\,\d \phi^2. 
\end{equation}
So, for instance, to write a circular polarisation we just have \index{gravitational waves!circular polarisation}
\begin{align}
 \d s^2 = \d u \, \d v &+ H(u) \, r^2\,\left\{ \cos(2\phi)\, \cos(\Omega u) \, + \sin(2\phi)\, \sin(\Omega u) \, \right\} \, \d u^2 \nonumber \\ 
 & + \d r^2 + r^2\,\d \phi^2. 
\end{align}

\vspace{20pt}

\noindent
In contrast the Rosen form cannot be separated in this simple manner, as that would require \index{gravitational waves!Rosen form}
\begin{equation}\label{pp:d}
R_{uu} = - \left\{ \frac{1}{2} \; g^{AB} \; g_{AB}'' 
- \frac{1}{4} \; g^{AB} \; g_{BC}' \; g^{CD} \; g_{DA}' \right\}
\end{equation}
to easily decouple in some way, and there is no obvious way in which this happens. This means there is no clear way to write down arbitrary polarisations in the Rosen form. Even simple, physically important scenarios such as circular or elliptical polarisation pose a difficulty.

This is only a problem for the strong-field case: due to superposition in the linearised case we can clearly decouple and add the different polarisations to create any arbitrary polarisation. 

Note again that from section (\ref{pp:e}) we have the same pattern of non-zero components in the Rosen and Brink\-mann forms (cf. equations (\ref{pp:f}), (\ref{pp:g}), (\ref{pp:h})), which indicates that \emph{some} of our intuition regarding polarisation modes will also carry over to the Rosen form of the metric. Since these two forms are related in this way, and polarisation modes are related to the non-zero components of the Riemann tensor, the comparative difficulty in constructing polarisations of the Rosen form is somewhat puzzling, and suggests that equation (\ref{pp:d}) should simplify in some way. Investigating this curiosity, a simplification of the vacuum equations and hence a way to write general polarisations is detailed below.

\section{Linear Polarisations}\index{gravitational waves!linear polarisation}

Consider the relatively simple strong-field gravitational wave metric in the $+$ linear polarisation \cite{D'Inverno}. That is, set $g_{xy} = 0$ so that $g_{AB}$ has only two nontrivial components, $g_{xx}$ and $g_{yy}$, corresponding to oscillations along the $x$ and $y$ axes. The resulting metric can be written in the form
\begin{equation}
\label{rosen:a}
\d s^2 = \d u\d v + f^2(u) \, \d x^2 + g^2(u)\,\d y^2.
\end{equation}
The only non-zero component of the Ricci tensor is: 
\begin{equation}
R_{uu} = - \left\{ \frac{f''}{ f} + \frac{g''}{g} \right\}.
\end{equation}
Although this expression for the Ricci tensor is compact, ultimately this form of the metric turns out not to be very tractable. However, if we write the metric in the form
\begin{equation}\label{rosen:b}
\d s^2 = \d u\d v + S^2(u) \, \left\{ e^{+X(u)} \,\d x^2 + e^{-X(u)} \, \d y^2 \right\},
\end{equation}
then
\begin{equation}
R_{uu} = - \frac{1}{2} \,\left\{ 4 \; \frac{S''}{S} + {(X')^2} \right\}.
\end{equation}
This may not initially look very promising, but it is this version of the metric that permits us to make the most progress. 
In particular note that in vacuum we have
\begin{equation}
X' = 2 \sqrt{-S''/S},
\end{equation}
so that
\begin{equation}
X(u) = 2 \int^u \sqrt{-S''/S} \d u.
\end{equation}
and the general vacuum wave for $+$ polarisation can be put in the form
\begin{eqnarray}\label{E:3}
\d s^2 &=& \d u\d v
+ S^2(u) \; \left\{ \exp\left( 2 \int^u \sqrt{-S''/S} \d u \right) \d x^2 \right.
\nonumber
\\
&&
\qquad\qquad\qquad\qquad \left. + \exp\left( -2 \int^u \sqrt{-S''/S} \d u \right) \d y^2 \right\}.
\quad
\end{eqnarray}
Note that as expected from the Brinkmann form we have one free function per polarisation mode.

\vspace{20pt}

\noindent
To take the strong-field gravitational wave metric in the $\times$ linear polarisation consider
\begin{equation}
\d s^2 =\d u\d v + \frac{f^2(u)+g^2(u)}{2}[ \d x^2 + \d y^2 ] + [f^2(u)-g^2(u)] \d x \d y,
\end{equation}
which could also be obtained from equation (\ref{rosen:a}) simply by performing a 45 degree rotation in the $x$--$y$ plane.
The only non-zero component of the Ricci tensor is again: 
\begin{equation}
R_{uu} = - \left\{ \frac{f''}{f} + \frac{g''}{g} \right\}.
\end{equation}
If we now write this metric in the form
\begin{equation}
\label{rosen:c}
\d s^2 = \d u\d v + S^2(u) \left\{ \cosh(X(u)) [\d x^2 + \d y^2] + 2 \sinh(X(u)) \d x \d y \right\},
\end{equation}
(equivalent to rotating equation (\ref{rosen:b}) by $45^\circ$) then, as for the $+$ mode, we have
\begin{equation}
R_{uu} = - \frac{1}{2} \left\{ 4 \; \frac{S''}{S} + {(X')^2} \right\}.
\end{equation}
In vacuum we can again solve for $X(u)$ and now obtain
\begin{eqnarray}
\d s^2 &=& \d u\d v 
+ S^2(u) \left\{ \cosh\left( 2 \int^u \sqrt{-S''/S} \d u \right) [\d x^2 + \d y^2] \right.
\nonumber\\
&&
\qquad \qquad \qquad \left. + 2 \sinh\left( 2 \int^u \sqrt{-S''/S} \d u \right) \d x \, \d y \right\}.
\end{eqnarray}
(as above, we could have obtained this form by rotating equation (\ref{E:3}) by $45^\circ$). There is again one freely specifiable function for this $\times$ linear polarisation mode. 

By rotating the $x$--$y$ plane through a fixed but arbitrary angle $\Theta_0$ we can easily deal with linear polarisation modes along any desired axis. Thus there is no difficulty in representing the $+$, $\times$ or any \emph{linear} polarisation in the Rosen form. The difficulty only arises when considering $u$-dependent polarisations. 


\section{Arbitrary polarisation}


Let us now take an arbitrary, possibly $u$ dependent, polarisation and consider the following metric ansatz, which was found through repeated experimentation:
\begin{align}
\d s^2 &= \d u\d v \nonumber\\
&+ S^2(u) \; \left\{ \cosh(B(u)) [e^{+X(u)}\,d x^2 + e^{-X(u)}\,d y^2] + 2 \sinh(B(u))\, \d x \d y \right\}.
\end{align}
Note that this reduces to equation (\ref{rosen:b}) when $B(u)=0$, corresponding to $+$ polarisation, while setting $X(u)=0$ (and changing the arbitrary function's name from $B(u)$ to $X(u)$) gives equation \eqref{rosen:c} corresponding to $\times$ polarisation. Furthermore we have sufficient free functions, namely \{$S(u)$, $B(u)$, $X(u)$\}, to completely saturate the arbitrary $2\times 2$ symmetric matrix $g_{AB}(u)$. A brief calculation yields the only nonzero component of the Ricci tensor:
\begin{equation}
R_{uu} = - \frac{1}{2} \left\{ 4 \, \frac{S''}{S} + {(B')^2} + \cosh^2[B(u)] \; (X')^2 \right\}.
\end{equation}
Note that $B(u)$ and $X(u)$ have \emph{not} decoupled --- however a hint on how to proceed is provided by noting that the 2-metric,
\begin{equation}
\d B^2 + \cosh^2 (B) \d X^2,
\end{equation}
is one of many possible ways of representing the metric of the hyperbolic plane, $H_2$.

Let us try a slightly different representation of the same general metric:
\begin{eqnarray}
\d s^2 &=& \d u\d v + S^2(u) \; \left\{ \vphantom{\Big|} [ \cosh(X(u)) + \cos(\theta(u)) \sinh(X(u))] \d x^2 \right.
\nonumber
\\
&&
\qquad \left. \vphantom{\Big|} + 2 \sin(\theta(u)) \sinh(X(u)) \d x \d y \right.
\nonumber
\\
&&
\qquad \left. \vphantom{\Big|} + [\cosh(X(u)) - \cos(\theta(u)) \sinh(X(u))] \d y^2 \right\}.
\end{eqnarray}
Note that setting $\theta(u)=0$ corresponds to $+$ polarisation, whilst setting $\displaystyle \theta(u)=\pi/2$ corresponds to $\times$ polarisation, and in general $ \theta(u) = \Theta_0/2$ corresponds to linear polarisation along axes rotated by an angle $\Theta_0$. 

Thus this form of the metric introduces a simple physical interpretation for the functions, unlike the previous metric. Here, $\theta(u)$ describes the angular dependence while $S(u)$ can be regarded as an ``envelope'' function, describing the strength of the wave, and $X(u)$ is the shape of the ``distortion''. Furthermore we again have sufficient free functions, now \{$S(u)$, $X(u)$, $\theta(u)$\}, to completely saturate the arbitrary $2\times 2$ symmetric matrix $g_{AB}(u)$.

Now
\begin{equation}
R_{uu} = - \frac{1}{2} \left\{ 4 \; \frac{S''}{S} + {(X')^2} + \sinh^2(X(u)) \; (\theta')^2 \right\},
\end{equation}
and the vacuum equations are simply
\begin{equation}
 4 \, \frac{S''}{S} + {(X')^2} + \sinh^2(X(u)) \; (\theta')^2 = 0.
\end{equation}
Let us introduce a dummy function, $L(u)$, and split this into the two equations
\begin{equation}\label{firstarb}
 4\, \frac{S''}{S} + {(L')^2} = 0,
\end{equation}
and
\begin{equation}\label{secondarb}
(L')^2 = {(X')^2} + \sinh^2(X(u)) \; (\theta')^2.
\end{equation}
The first of these, equation (\ref{firstarb}), is simply the equation you would have to solve for a pure $+$ or $\times$ or in fact any linear polarisation.
The second of these, equation (\ref{secondarb}), can be rewritten as
\begin{equation}
\d L^2 = \d X^2 + \sinh^2(X) \d\theta^2,
\end{equation}
and is just the statement that $L$ can be interpreted as distance in the two dimensional hyperbolic plane, $H_2$.

\subsection{Algorithm}
\label{algorithm}
This now provides us with a very straightforward algorithm for constructing any arbitrary polarisation strong-field gravitational wave in the Rosen form:
\begin{itemize}
\item Pick an arbitrary $L(u)$ and solve:
\begin{equation}
 4 \; \frac{S''}{S} + {(L')^2} = 0.
\end{equation}
\item 
Pick an arbitrary curve in the $(X,\theta)$ plane such that $L(u)$ is hyperbolic arc-length along that curve:
\begin{equation}
\d L^2 = \d X^2 + \sinh^2(X) \d\theta^2.
\end{equation}
(alternatively one could consider these two steps in reverse.)
\item
This construction then solves the vacuum Einstein equations for the metric
\begin{eqnarray}
\d s^2 &=& \d u\d v + S^2(u) \left\{ \vphantom{\Big|} [ \cosh(X(u)) + \cos(\theta(u)) \sinh(X(u))] \d x^2 \right.
\nonumber
\\
&&
\qquad \left. \vphantom{\Big|} + 2 \sin(\theta(u)) \sinh(X(u)) \d x \d y \right.
\nonumber
\\
&&
\qquad \left. \vphantom{\Big|} + [\cosh(X(u)) - \cos(\theta(u)) \sinh(X(u))] \d y^2 \right\}.
\end{eqnarray}
\end{itemize}
In this sense we have completely solved arbitrary polarisation strong-field gravitational waves in the Rosen form.

\subsection{Comparison to electromagnetism}
Note the similarities (and differences) with regard to the familiar Maxwell electromagnetism (and with respect to the Brinkmann form). In electromagnetism the two independent linear polarisations can be specified by 
\begin{equation}
\vec E(u) = E_x(u) \; \hat x + E_y(u)\; \hat y,
\end{equation}
with no additional constraints (compare with equation (\ref{pp:b})). Thus an electromagnetic wavepacket of arbitrary polarisation can be viewed as an arbitrary ``walk'' in the $(E_x, E_y)$ plane. 

We could also go to a magnitude-phase representation $(E,\theta)$ where
\begin{equation}
\vec E(u) = E(u) \cos\theta(u) \; \hat x + E(u) \sin\theta(u)\; \hat y.
\end{equation}
(Compare this with equation (\ref{pp:c})). So an electromagnetic wavepacket of an arbitrary polarisation can also be viewed as an arbitrary ``walk'' in the $(E, \theta)$ plane, where the $(E, \theta)$ plane is provided with the natural Euclidean metric
\begin{equation}
\d L^2 = \d E^2 + E^2 \d \theta^2.
\end{equation}
In contrast, for gravitational waves in the Rosen form we are now dealing with an arbitrary ``walk'' in the hyperbolic plane, $H_2$, rather than in the Euclidean plane. Furthermore, because of the nonlinearity of strong-field general relativity there is still one remaining differential equation to solve for the ``envelope'', $S(u)$.

\section{Circular polarisation}\index{gravitational waves!circular polarisation}

We can now adopt the above discussion to formulate strong-field circular polarisation in the Rosen form. We emphasise again that there is no difficulty whatsoever with weak-field, linearised circular polarisation, it is only for strong fields that it is difficult to formulate circular polarisation in the Rosen form. Circular polarisation corresponds to
\begin{equation}
\theta(u) = \Omega_0\; u; \qquad X(u) = X_0.
\end{equation}
That is, a fixed distortion $X_0$ with the plane of polarisation advancing linearly with retarded time $u$ (with a coordinate choice to set the constant term to zero).   The envelope function, $S(u)$, must be determined through the Einstein equations. Now the metric becomes
\begin{eqnarray}
\d s^2 &=&\d u\d v + S^2(u) \; \left\{ \vphantom{\Big|} [ \cosh(X_0) + \cos(\Omega_0\; u) \sinh(X_0)] \d x^2 \right.
\nonumber
\\
&&
\qquad \left. \vphantom{\Big|} + 2 \sin(\Omega_0 u) \sinh(X_0) \d x \d y \right.
\nonumber
\\
&&
\qquad \left. \vphantom{\Big|} + [\cosh(X_0) - \cos(\Omega_0 u) \sinh(X_0)] \d y^2 \right\}. 
\end{eqnarray}
The only nontrivial component of the Ricci tensor is then
\begin{equation}
R_{uu} = - \frac{1}{2} \left\{ 4 \; \frac{S''}{S} + \sinh^2(X_0) \; \Omega_0^2 \right\},
\end{equation}
and the vacuum equations are simply
\begin{equation}
 S'' = -\frac{ \sinh^2(X_0) \; \Omega_0^2}{4} \; S.
\end{equation}
To explicitly link this to the algorithmic construction given in section (\ref{algorithm}), note that $\Omega_0 \sinh(X_{0})$ is the hyperbolic arclength of a circle segment. Now we can solve for $S(u)$,
\begin{equation}
S(u) = S_0 \; \cos\left\{ \frac{ \sinh(X_0) \; \Omega_0 \; (u-u_0)}{2}\right\}.
\end{equation}

So explicitly,
\begin{eqnarray}
\d s^2 &=& \d u\d v +S_0^2 \; \left(\cos\left\{\frac { \sinh(X_0) \; \Omega_0 \; (u-u_0)}{2}\right\}\right)^2 \; 
\nonumber
\\
&&
\qquad \left\{ \vphantom{\Big|} [ \cosh(X_0) + \cos(\Omega_0\; u) \sinh(X_0)] \d x^2 \right.
\nonumber
\\
&&
\qquad \left. \vphantom{\Big|} + 2 \sin(\Omega_0 u) \sinh(X_0) \d x \d y \right.
\nonumber
\\
&&
\qquad \left. \vphantom{\Big|} + [\cosh(X_0) - \cos(\Omega_0 u) \sinh(X_0)] \d y^2 \right\}. 
\end{eqnarray}
This now describes a spacetime that has good reason to be called a strong-field circularly polarized gravitational wave.

\vspace{20pt}

\noindent
For consistency, note that the weak-field limit \index{gravitational waves!weak-field}\index{gravitational waves!circular polarisation}corresponds to $X_0\ll 1$ so that for an arbitrarily long interval in retarded time $u$ we have $S \approx S_0$, and without loss of generality we can set $S\approx 1$. Then
\begin{align}\label{weakapprox}
\d s^2 \approx &\; \d u\d v + \d x^2 + \d y^2 \nonumber \\ 
+& X_0 \; \left\{ \vphantom{\Big|} \cos(\Omega_0\; u) [\d x^2-\d y^2] + 2 \sin(\Omega_0\; u) \d x \d y \right\},
\end{align}
and compare the above (equation (\ref{weakapprox})) to equation (\ref{weak:circular}).

Further generalisations to elliptic polarisation are tedious, but given the significantly more general algorithm of the preceding subsection, reasonably straightforward. 


\section{Decoupling the general form in arbitrary dimensions}\label{rosen:decompose}\index{unimodular decomposition}

Considering the calculations considered so far, note we have made progress with these forms by separating out a unit determinant piece and some envelope function. Based on this, one might suspect that there is some general decoupling between the overall ``envelope'' of the gravitational wave and the ``directions of oscillation''. Let us return to considering the metric in general Rosen form
\begin{equation}
\d s^2 = \d u\d v + g_{AB}(u)\d x^A\d x^B, 
\end{equation}
where for generality the $x^A, x^B$ are now any arbitrary number of dimensions ($d_\perp\geq 2$) transverse to the $(u,v)$ plane. 
It is easy to check that the only non-zero component of the Ricci tensor is still
\begin{equation}
R_{uu} = - \left\{ \frac{1}{2} \; g^{AB} \; g_{AB}'' 
- \frac{1}{4} \; g^{AB} \; g_{BC}' \; g^{CD} \; g_{DA}' \right\}.
\end{equation}
Let us now decompose the $d_\perp \times d_\perp$ matrix $g_{AB}$ into an ``envelope'' $S(u)$ and a unit determinant metric related to the ``direction of oscillation''. That is, let us take
\begin{equation}
g_{AB}(u) = S^2(u)\; \hat g_{AB}(u),
\end{equation}
where $\det( \hat g) \equiv 1$. A related discussion can be found in section 109 of Landau and Lifshitz \cite{Landau}. Our goal is to see if we can make the overall ``envelope'' $S(u)$ decouple from $\hat g_{AB}(u)$. 
To start, note that
\begin{equation}
g_{AB}' = 2 S\; S' \; \hat g_{AB} + S^2\; \hat g_{AB}',
\end{equation}
and
\begin{equation}
g_{AB}'' = 2 S\; S'' \; \hat g_{AB} + 2 S'\; S' \; \hat g_{AB} + 4 S\; S' \; \hat g_{AB}' + S^2\; \hat g_{AB}''.
\end{equation}
Therefore
\begin{eqnarray}\label{secdervrosen}
g^{AB} \; g_{AB}'' &=& 
 \frac {1}{S^2} \left\{ 2 (S S'' + S' S') d_\perp + 4 S\; S' \; [ \hat g^{AB} \; \hat g_{AB}'] + S^2\; [ \hat g^{AB} \; \hat g_{AB}''] \right\}
\nonumber\\
&=&
 2 \left(\frac{S''}{S} + \frac{S' S'}{S^2}\right) d_\perp + 4 \frac{S'}{S} \; [ \hat g^{AB} \; \hat g_{AB}'] + [ \hat g^{AB} \; \hat g_{AB}''],
 \end{eqnarray}
 and similarly
\begin{align}\label{firstdervrosen}
&g^{AB} \; g_{BC}' \; g^{CD} \; g_{DA}' \nonumber \\
&= g^{AB} [ 2 S\; S' \; \hat g_{BC} + S^2\; \hat g_{BC}' ] g^{CD} [2 S\; S' \; \hat g_{DA} + S^2\; \hat g_{DA}'] \nonumber\\
&= \frac{1}{S^4} \left\{ 4 (S\; S')^2 d_\perp + 4 S^3 S' [ \hat g^{AB} \; \hat g_{AB}'] + S^4 [ \hat g^{AB} \; \hat g_{BC}' \; \hat g^{CD} \; \hat g_{DA}' ]\right\}\nonumber\\
&= 4 \left(\frac{S'}{S}\right)^2 d_\perp + 4\left(\frac{S'}{S}\right) [ \hat g^{AB} \; \hat g_{AB}'] + [ \hat g^{AB} \; \hat g_{BC}' \; \hat g^{CD} \; \hat g_{DA}' ].
 \end{align}
Now combine the results of equations (\ref{firstdervrosen}) and (\ref{secdervrosen})
\begin{align}
R_{uu} &= 
- \left\{ \frac{1}{2} \; g^{AB} \; g_{AB}'' 
- \frac{1}{4} \; g^{AB} \; g_{BC}' \; g^{CD} \; g_{DA}' \right\}\nonumber\\
&=- \left(\frac{S''}{S} + \frac{S' S'}{S^2}\right) d_\perp -2 \frac{S'}{S} \; [ \hat g^{AB} \; \hat g_{AB}'] - \frac{1}{2} [ \hat g^{AB} \; \hat g_{AB}''] \nonumber\\
& + \left(\frac{S'}{S}\right)^2 d_\perp + \left(\frac{S'}{S}\right) [ \hat g^{AB} \; \hat g_{AB}'] + 
 \frac{1}{4} [ \hat g^{AB} \; \hat g_{BC}' \; \hat g^{CD} \; \hat g_{DA}' ]\nonumber\\
&= - \frac{S''}{S} d_\perp - \frac{1}{2} [ \hat g^{AB} \; \hat g_{AB}''] \nonumber \\
&+ \frac{1}{4} [ \hat g^{AB} \; \hat g_{BC}' \; \hat g^{CD} \; \hat g_{DA}' ]
 - \left(\frac{S'}{S}\right) [ \hat g^{AB} \; \hat g_{AB}']. 
\end{align}
Because we have defined that $\det( \hat g) \equiv 1$, we have as a matrix identity
 \begin{equation}
 [ \hat g^{AB} \; \hat g_{AB}'] = 0,
 \end{equation}
 and differentiating this one more time
 \begin{equation}
 [ \hat g^{AB} \; \hat g_{AB}''] - [ \hat g^{AB} \; \hat g_{BC}' \; \hat g^{CD} \; \hat g_{DA}' ] = 0.
 \end{equation}
Therefore 
\begin{equation}\label{rosen:d}
R_{uu} = - d_\perp \, \frac{S''}{S} - \frac{1}{4}\, [ \hat g^{AB} \; \hat g_{BC}' \; \hat g^{CD} \; \hat g_{DA}' ],
\end{equation}
or more abstractly,
\begin{equation}
R_{uu} = - d_\perp \, \frac{S''}{S} - \frac{1}{4} \tr \left\{[\hat g]^{-1} \; [\hat g]' \; [\hat g]^{-1} \; [\hat g]' \right\}. 
\end{equation}
Note that we have now succeeded in decoupling the determinant ($\displaystyle \det(g) = S^{2d_\perp}$; effectively the ``envelope'' $S(u)$ of the gravitational wave) from the unit-determinant matrix $\displaystyle \hat g(u)$. This observation is compatible with all the specific examples considered above.

\vspace{20pt}

\noindent
Now, analogously to section (\ref{algorithm}), consider the set $\displaystyle SS(I\!\!R, d_\perp)$ of all unit determinant real symmetric matrices, and on that set (not a group) consider the Riemannian metric,
\begin{equation}
\d L^2 = \tr\left\{ [\hat g]^{-1} \d[\hat g] [\hat g]^{-1} \d[\hat g] \right\}.
\end{equation}
Then
\begin{equation}
R_{uu} = - \frac{1}{4} \left\{ 4 d_\perp\, \frac{S''(u)}{S(u)} + \left(\frac{d L}{d u}\right)^2 \right\}. 
\end{equation}
The vacuum Einstein equations then reduce to
\begin{equation}
\frac{d L}{d u} = 2 \,\sqrt{ -  d_\perp \; \frac{S''}{S}}; \qquad L(u) = \int^u 2\,\sqrt{ - d_\perp \; \frac{S''}{S}} d u.
\end{equation}
That is, an arbitrary polarisation vacuum Rosen wave is an arbitrary walk in $\displaystyle SS(I\!\!R, d_\perp)$, with distance along the walk $L(u)$ being related to the envelope function $S(u)$ in the differential relation above.

\vspace{20pt}
\noindent
To probe the polarisation modes in more detail, a completely analogous but slightly more complicated calculation determines the Riemann tensor components to be\index{Riemann tensor}
\begin{eqnarray}
R_{uAuB} = - \left\{
 S S'' \hat g_{AB} +S^2 \left[ \frac{1}{2} \; \hat g''_{AB} -\frac{1}{4} \; \hat g_{AC}' \; \hat g^{CD} \; \hat g_{DB}' \right] + S S' \hat g'_{AB}
 \right\}.
\end{eqnarray}
Note that this is compatible with equation (\ref{pp:e}).

\section{Discussion}

The {\it pp} waves, and gravitational plane waves in particular, are interesting for a number of reasons, including such diverse topics as the study of colliding plane waves and quantum theories of gravity. 

In spite of gravitational waves in both the Rosen and the Brinkmann forms having been known for many decades they are still of considerable interest in research and, as we have found, there are still a number of surprises in the Rosen form.

The two forms that are commonly used for gravitational waves have their respective advantages and disadvantages. However, the fact that it is entirely trivial to represent arbitrary polarisations in Brinkmann form, but unclear how to do so in the Rosen form, hints that the previous understanding of polarisations of the Rosen form was incomplete.

We have found that splitting the metric into an ``envelope'' function and a unit determinant matrix succeeds in decoupling the vacuum equations into parts depending only on a walk in some appropriately defined polarisation space, and a further differential equation relating this walk to the ``envelope'' function. In (3+1) dimensions this polarisation space is the hyperbolic plane. 

This decoupling, and the corresponding defining of the polarisation space, can be extended to an arbitrary number of dimensions. Furthermore we have applied this construction to a case of particular interest, the circularly polarized wave. 

\chapter[Penrose limits, ultra-local limits and Bianchi cosmologies]{Penrose limits, ultra-local limits and Bianchi cosmologies}\label{penrose}

In this chapter we will examine several seemingly unrelated aspects of general relativity. These topics will all be relevant for understanding the next chapter, in which knowledge of all the subjects presented will be useful. 

Firstly, we will look at approximations around an arbitrary point and an arbitrary null geodesic in any spacetime, the latter of which is known as a Penrose limit, and produces the {\it pp} wave spacetimes examined in the previous chapter.

Next we will move on to various cosmological models, starting with the simple FLRW spacetime, then the set of all homogenous cosmologies. We do this not because cosmological concerns are directly important to the research presented, but because this is the natural context in which the spacetimes useful for the research which follows in the next chapter are most frequently encountered.

We will also look at other sorts of limits: ultra-local limits, which have mostly been developed in a quantum gravity context, and some simplifications possible in cosmological contexts looking at approximations near singularities. 


\section{The flat space limit}\label{flat}\index{flat space limit|(}

It is clear that the zeroth term in an expansion for a general spacetime will be flat spacetime, so the limit for an arbitrary spacetime around a general point will be a particular flat space limit. We can precisely describe this limit as follows:

\vspace{20pt}
\noindent
Begin with an arbitrary spacetime
\begin{equation}
\d s^2=g_{ab}(\vec x)\,\d x^a\, \d x^b,
\end{equation}
and implement the limit in stages:
\begin{enumerate}
\item Change coordinates
\begin{equation}
x^a \to \lambda\, x^a,
\end{equation}
to obtain
\begin{equation}
\d s^2=\lambda^2\,g_{ab}(\lambda\, \vec x)\,\d x^a\, \d x^b.
\end{equation}
\item Perform a conformal deformation
\begin{equation}
\d \bar s^2=\frac{\d s^2}{\lambda^2}=g_{ab}(\lambda\, \vec x)\,\d x^a\, \d x^b.
\end{equation}
\item Now take the limit as $\lambda \to 0$
\begin{equation}
\d \bar s^2=g_{ab}(0)\,\d x^a\, \d x^b.
\end{equation}
\end{enumerate}
This construction gives us the flat space approximation to the metric around the point $\vec x =0$. This could then be used as the zeroth term of an expansion to approximate the spacetime close to $\vec x=0$.

Such an expansion is already well-known, using a Riemann normal coordinate system expansion. Riemann normal coordinates are coordinates where \emph{at a point} (chosen to be $\vec x =0$) the metric is $\eta_{ab}$ and the Christoffel symbols\index{Christoffel symbol} are zero. Such a coordinate choice is always possible. Then the expansion about $\vec x = 0$ is\index{Riemann normal coordinates}
\begin{equation}
g_{ab}(x)=\eta_{ab}+\frac{1}{3}\,R_{acdb}\,x^c\,x^d +\frac{1}{6}\nabla_e\,R_{acdb}\,x^c\,x^d\, x^e +O(x^4).
\end{equation}
For instance, see \cite{Muller}.\index{flat space limit|)}


\section{Penrose limits}\label{penroselim}\index{Penrose limit|(}

The previous chapter dealt with plane waves. Although some of the applications and areas of interest for these spacetimes were mentioned, one context in which these spacetimes have been studied was omitted: that they are the result of Penrose limits. This limit, suggested by Roger Penrose in 1976 \cite{Penrose2}, is the zeroth approximation for a spacetime around an arbitrary null geodesic.

\subsection{Construction}\label{penrose:construction}

To construct the Penrose limit pick an arbitrary null geodesic, $\gamma$, in any arbitrary spacetime, and in some tube surrounding the geodesic choose a set of adapted coordinates\index{geodesic!null}, also sometimes referred to as Penrose coordinates,
\begin{equation}\label{penrose:a}
\d s^2_\gamma =  \d U\, \d V +a(U, V, y^i)\,\d V^2+b_i(U, V, y^i)\,\d V\, \d y^i +g_{ij}(U, V, y^i)\,\d y^i\, \d y^j.
\end{equation}
It is always possible to put a spacetime into these coordinates locally \cite{blaulecture2} (this is not required globally and as such does not impose any restriction on the type of spacetime considered).\index{adapted coordinates} This coordinate system corresponds to embedding a null geodesic in a congruence of twist-free null geodesics, given by $V$ and $y^i$ constant.

\vspace{20pt}
\noindent
Now implement the Penrose limit in stages:
\begin{enumerate}
\item
Make a coordinate transform
\begin{equation}\label{penrose:trans}
V = \lambda^2\, v ; \qquad y^i = \lambda\, x^i; \qquad U = u,
\end{equation}
then
\begin{align}
\d s^2_\gamma = \lambda^2\, \d u\, \d v &+\lambda^4\,a(u, \lambda^2v, \lambda y^i)\,\d v^2+\lambda^3\,b_i(u, \lambda^2v, \lambda y^i)\,\d v\, \d y^i \nonumber \\
&+\lambda^2\,g_{ij}(u, \lambda^2v, \lambda y^i)\,\d y^i\, \d y^j.
\end{align}

\item
Perform a conformal deformation
\begin{align}\label{deform}
\d \bar s^2_\gamma=\frac{\d s^2_\gamma}{\lambda^2} = \d u\, \d v &+\lambda^2\,a(u, \lambda^2v, \lambda y^i)\,\d v^2+2\lambda b_i(u, \lambda^2v, \lambda y^i)\,\d v\, \d y^i \nonumber \\
&+g_{ij}(u, \lambda^2v, \lambda y^i)\,\d y^i\, \d y^j.
\end{align} 

\item
Take the limit $\lambda \to 0$. The net effect is
\begin{equation}\label{penrose:lim}
\d \bar s^2 = \d u\, \d v +g_{ij}(u, 0, 0)\,\d x^i\, \d x^j.
\end{equation}
\end{enumerate}

Thus we see that the result of this limit is a {\it pp} wave spacetime\index{pp@\textit{pp} spacetime}. This could then be used as a zeroth approximation to the spacetime around the geodesic in some expansion.

Therefore this limit takes any lightlike geodesic in any spacetime and essentially ``blows up'' that geodesic to becomes what turns out to be a {\it pp} wave spacetime. A physical interpretation is that this is the limit of an accelerating observer approaching a lightlike geodesic whilst simultaneously adjusting their clocks to run progressively faster (ensuring the spacetime does not become degenerate), approaching the limit of measuring $u$ \cite{Penrose2}. 


Penrose limits have generated interest for a number of reasons. They are particularly important in the context of string theory where {\it pp} waves often provide exactly solvable backgrounds \cite{Gueven, Tseytlin}. Some cases that have been particularly well-studied are the limits of AdS${}_n\times$S${}^n$ and Schwarzschild or more general singularities \cite{BMN, blaulecture2, Blau2, Blau1}.

\subsection{Features of the Penrose limit}\label{penrose:hereditary}
\index{hereditary properties}
An important question for any limit is: what properties of the original metric are preserved in the particular limit? These properties are called hereditary properties \cite{blaulecture2, Geroch}. 

In the case of Penrose limits a property will be hereditary when it remains the same through the various steps in the limit (\ref{penrose:trans}) -- (\ref{penrose:lim}). In many cases we are most interested in coordinate independent properties, and in this case we only need to consider whether such properties survive rescaling (\ref{deform}) and the limiting process (\ref{penrose:lim}). 

One important result is that if there is a tensor field which is zero for all $\lambda > 0$, then this field will also be zero in the limit \cite{Geroch}. Thus, for instance, a Ricci flat spacetime will have a Ricci flat Penrose limit. 

Penrose limits do not preserve all structures that may be of interest. For instance they erase the presence of horizons \cite{Hubeny, Senovilla}. Note that Penrose limits do \emph{not} preserve the form of the stress-energy. We previously noted that {\it pp} waves have only one non-zero component of the Ricci and Einstein tensors, allowing only vacuum or null dust. This means that the Penrose limit of any spacetime also only has at most one component of the stress-energy, $T_{uu}$ \cite{ blaulecture2, Penrose1, Penrose2, steele}.\index{stress-energy tensor}\index{pp@\textit{pp} spacetime}\index{Penrose limit|)}


\section{FLRW cosmology}\label{penrose:FLRW}\index{FLRW metric|(}\index{cosmology}


Our fundamental understanding of cosmology is based on the evidence that the universe is highly homogeneous and isotropic (and has has been back to at least the time of the decoupling of the CMB). This gives us the information that (averaging over sufficiently large volumes) the matter in the universe is well described by a perfect fluid, and 
\begin{equation}
R_{ij}=\frac{1}{3}\,R\,g_{ij},
\end{equation}
implying that the curvature of space is constant, leaving essentially only three options of flat, spherical or hyperbolic space. 

The direct consequence of these symmetries is that we have a very simple spacetime: the Friedmann-Lemaitre-Robertson-Walker (FLRW) spacetime:
\begin{equation}
\d s^2 = -\d t^2 + a(t)^2\,\left\{ \frac{\d r^2}{1-kr^2}+r^2\, \left[ \d \theta^2+\sin^2(\theta)\, \d \phi^2 \right] \right\},
\end{equation}
where the free parameter, $a(t)$, is a scale factor describing the ``size'' of the universe\index{scale factor}, and 
\begin{itemize}
\item $k\,=\,1 \Rightarrow$ positive spatial curvature, spatial slices have spherical geo\-metry.
\item $k\,=\; 0 \Rightarrow$ zero spatial curvature, spatial slices are flat.
\item $k\,=\,-1 \Rightarrow$ negative spatial curvature, spatial slices have hyperbolic geometry.
\end{itemize}

\subsection{Einstein equations and conservation law}\label{flrwee}
The Einstein equations produce these relations, describing the evolution of the universe in terms of the matter present:\index{Einstein equations}
\begin{equation}\label{flrwrho}
8\pi\, \rho = 3\left[\frac{\dot a^2}{a^2}+\frac{k}{a^2}\right],
\end{equation}
\begin{equation}\label{penrose:c}
8\pi\, p = -\left[\frac{\dot a^2}{a^2}+\frac{k}{a^2}+ 2\frac{\ddot a}{a}\right].
\end{equation}
It is sometimes useful to rearrange the Einstein equations as:
\begin{equation}
R_{ab}=T_{ab} - \frac{1}{2}\,T\, g_{ab},
\end{equation}
which lets us rewrite the above relations (equations (\ref{flrwrho}) and (\ref{penrose:c})) as
\begin{equation}\label{penrose:j}
\frac{4}{3}\,\pi \left[\rho+3p\right] =-\frac{\ddot a}{a},
\end{equation}
and
\begin{equation}\label{penrose:k}
\frac{\ddot a}{a} + 2 \, \frac{\dot a^2}{ a^2} +2\frac{k}{a^2}= 4\pi(\rho-p).
\end{equation}
Furthermore, if we apply covariant conservation\index{conservation law} to this spacetime geometry we obtain a conservation law,
\begin{equation}\label{penrose:d}
\frac{\d}{\d t}\left[ \rho\; a^3\right] = -p\frac{\d}{\d t}[a^3].
\end{equation}

\subsection{Raychaudhuri equation}
\index{Raychaudhuri equation}

We can simplify the (timelike) Raychaudhuri equation (as described in section (\ref{raychaud})) by noticing that we have a natural timelike congruence defined by $u = \partial_t$, while we also have $u^a = (1,0,0,0)$. For this timelike congruence the vorticity, shear and acceleration are zero. 
Furthermore, \index{expansion}
\begin{equation}
\theta = -3\, \frac{\ddot a}{a}.
\end{equation}

Therefore we can simplify the Raychaudhuri equation to:
\begin{equation}\label{FLRW:ray}
R_{ab}\,u^a\,u^b=-3\, \frac{\ddot a}{a}.
\end{equation}
In this simple spacetime the Raychaudhuri equation is reduced to a statement about the Ricci tensor:
\begin{equation}
R_{tt}= -3\, \frac{\ddot a}{a}.
\end{equation}
\index{FLRW metric|)}


\section{Bianchi cosmologies}\index{Bianchi classification}\index{cosmology}
Although the universe we see today is highly homogenous and isotropic an interesting question is whether it was always this way, or if it was possible to start in a highly inhomogeneous or anisotropic state and then, through the action of the Einstein equations and dissipative physics, produce the highly isotropic state we see today. Could a large range of initial conditions lead to a universe like ours? Another question that made people look outside the FLRW metric is the horizon problem, whereby far distant parts of the sky are mysteriously in thermal equilibrium. While some of these problems are now better understood in terms of an inflationary era, for a time one approach was to investigate these issues by using anisotropic (or inhomogeneous) classical general relativity, and some of those results are still of considerable interest. 
	
Due to the need for mathematical simplicity one of the ways this is investigated is to initially retain the assumption of homogeneity while allowing anisotropy. This may seem like quite a severe physical restriction, but some of the results in homogenous spacetimes have since been extended to inhomogenous ones. 

Homogeneous spacetimes are given by the nine types in the Bianchi classification \cite{Bianchi, EM}. This classification was known before the development of general relativity via a study of Lie algebras on homogenous 3D topologies. The Bianchi cosmologies are found by adding on a time dependence to the 3D metric of the classification, and embedding this in four dimensions, so that the Bianchi cosmologies can be described by the metric \cite{Podolskyexact, kramerexact}
\begin{equation}
\d s^2 = -\d t^2 +g_{kl}(t)\,\left(e^k{ }_i\,\d x^i\right)\left(e^l{ }_j\,\d x^j\right),
\end{equation}
where the basis vectors $e^a$ are functions of the spatial variables and are related by the Lie bracket
\begin{equation}
[e_k, e_l] = C^m{}_{kl}\,e_m,
\end{equation}
with $C^m{}_{kl}$ known as the structure constants\index{structure constants} which are, by construction, anti\-symmetric. The Bianchi types are then classified according to the form of the structure constants, which can be decomposed as 
\begin{equation}
C^m{}_{kl} = \epsilon_{klp}\,n^{pm}+2\delta^m{}_{[l}\,a_{k]},
\end{equation}
with $\epsilon_{klp}$ the Levi-Civita symbol, $n$ a diagonal tensor $\left(n_1, n_2, n_3\right) $ and $a_a=\left(a, 0, 0\right)$. Alternatively, the closely-related objects
\begin{equation}
C^{ab}=\epsilon^{dea}C^{b}{}_{de},
\end{equation}
are sometimes referred to as the structure constants. 

The Bianchi types are characterised by whether $a$ is non-zero and the sign of the components of $n$, as shown in table (\ref{bianchitypes}). See \cite{EM} for details.

\begin{table}[h!]\index{Bianchi classification}
\vspace{20pt}
\centering
\begin {tabular*}{0.75\textwidth} {@{\extracolsep{\fill}}ccccc}
\hline
Type 	&	a	&	 $n_{1}$		& $n_{2}$	& $n_{3}$	\\
\hline
I	&	0	&	0	&	0	&	0	\\
II 	&	0	&	+	&	0	&	0	\\
III	&	+	&	0	&	+	&	-	\\
IV	&	+	&	0	&	0	&	+	\\
V	&	+	&	0	&	0	&	0	\\
VI$_{0}$	&	0	&	+	&	-	&	0	\\
VI$_{a}$	&	a	&	0	&	+	&	-	\\
VII$_{0}$	&	0	&	+	&	+	&	0	\\
VII$_{a}$	&	a	&	0	&	+	&	+	\\
VIII 	&	0	&	+	&	+	&	-	\\
IX	&	0	&	+	&	+	&	+	\\
\hline
\end{tabular*}
\label{bianchitypes}
\caption{The Bianchi classification}
\vspace{25pt}
\end{table}

The Bianchi types are also often classified into class A with $a=0$ and class B with $a \neq 0$, so that class A consists of $\left\{I, II, VI_0, VII_0, VIII, IX\right\}$, and class B of $\left\{III, IV, V, VI_a, VII_a\right\}$. The familiar FLRW cosmology is a special case of several of the nine Bianchi types.

Furthermore it is possible to write down the Ricci tensor for generic Bianchi types as in \cite{Montani}\index{Ricci tensor},
\begin{equation}\label{bianchiR1}
R^0{}_0= -\frac{1}{2}\frac{\d g_{ij}}{\d t}g^{ji}-\frac{1}{4}\frac{\d g_{ij}}{\d t}g^{jk}\frac{\d g_{kl}}{\d t}g^{li};
\end{equation}
\begin{equation}\label{bianchiR2}
R^0{}_i = -\frac{1}{2}\frac{\d g_{kl}}{\d t}g^{lj}\left(C^k{}_{ji}-\delta^k{}_i C^{m}{}_{mj}\right);
\end{equation}
\begin{equation}\label{bianchiR3}
R^i{}_j = \frac{1}{2\sqrt{g}}\frac{\partial}{\partial t}\left(\sqrt{g} \frac{\d g_{jk}}{\d t}g^{ki} \right) - P^i{}_j.
\end{equation}
With
\begin{equation}
P_{ij} = -\frac{1}{2}\left(C^{kl}{}_j C_{kli}+C^{kl}{}_j C_{lki} -\frac{1}{2}C_j{}^{kl}C_{ikl}+C^k_{kl}C^l{}_{ij}+C^k{}_{kl}C_{ji}{}^l \right).
\end{equation}


\subsection{Vacuum Bianchi types}

Many studies are done in the case of the vacuum solutions of the Bianchi cosmologies, as under many circumstances of cosmological importance the overall behaviour is not affected by the form of matter. Looking at this case, it is common to use $e^\alpha = \left(l(x),\, m(x),\, n(x)\right)$, so
\begin{equation}
\d s^2 = -\d t^2 +\left(a^2(t)l_i(x)l_j(x)+b^2(t)m_i(x)m_j(x)+c^2(t)n_i(x)n_j(x)\right)\d x^i\d x^j.
\end{equation}
Then, simplifying the general results, 
\begin{align}
-R^0{}_0 &= \frac{\ddot a}{a}+ \frac{\ddot b}{b}+ \frac{\ddot c}{c}=0;\label{binachinine1} \\
-R^l{}_l &=\frac{\frac{d}{dt}(\dot a b c)}{abc}+\frac{1}{2a^2b^2c^2}\left(\lambda^2a^4-\left(\mu b^2-\nu c^2\right)\right)=0; \\
-R^m{}_m &=\frac{\frac{d}{dt}(a \dot b c)}{abc}+\frac{1}{2a^2b^2c^2}\left(\nu^2b^4-\left(\lambda a^2-\nu c^2\right)\right)=0; \\
-R^n{}_n &=\frac{\frac{d}{dt}(ab \dot c)}{abc}+\frac{1}{2a^2b^2c^2}\left(\nu^2c^4-\left(\mu b^2-\lambda a^2\right)\right)=0.\label{binachinine4}
\end{align}
Where $(\lambda, \mu, \nu)$ are the structure constants $(C_{ll}, C_{nn}, C_{mm})$, and the dot indicates a derivative with respect to $t$.\index{structure constants}

\vspace{20pt}

\noindent
Changing variables can also be helpful;
\begin{equation}
\alpha = \ln(a); \qquad \beta = \ln(b); \qquad \gamma = \ln(c),
\end{equation}
with a change of time coordinate
\begin{equation}
\d\tau =\frac{1}{abc}\,\d t.
\end{equation}
So the system of equations (\ref{binachinine1})--(\ref{binachinine4}) can now be described by
\begin{align}
2\ddot \alpha &= \left(\mu b^2 - \nu c^2\right)^2 - \lambda^2 a^4; \\
2\ddot \beta &= \left(\lambda a^2 - \nu c^2\right)^2 - \mu^2 b^4; \\
2\ddot \gamma &= \left(\lambda a^2 - \mu b^2\right)^2 - \nu^2 c^4;
\end{align}
and
\begin{equation}
\dot \alpha \dot \beta + \dot \alpha \dot \gamma + \dot \beta \dot \gamma = \frac{1}{4}\left(\lambda a^4+\mu^2 b^4 + \nu^2 c^4- 2\lambda\mu a^2 b^2- 2\lambda\nu a^2 c^2- 2\mu\nu b^2 c^2\right),
\end{equation}
where now the dot indicates a derivative with respect to $\tau$. 


\section{Bianchi Type I}\index{Bianchi type I}\index{Bianchi type I!diagonal}
The Bianchi type I is described by all the structure constants being zero. This means that we can always pick a triad $e^k{ }_i$ such that the three dimensional slice is given by Euclidean space. However, we want independence of such a coordinate choice when embedding in four dimensions, and so the general Bianchi type I is:
\begin{equation}
\d s^2 = -\d t^2 + g_{ij}(t) \; \d x^i \d x^j.
\end{equation}
Note the Bianchi type I includes FLRW with $k=0$ as a special case\index{FLRW metric}.

This is the simplest of the nine forms, and one of the most frequently studied (for a selection of classic and recent sources see \cite{Calogero1, Calogero2, Collins, Podolskyexact, Uggla, Misner:1967, kramerexact, Uzan1, Uzan2, Ryan}). 

In the cosmological context, it is very common to see the full Bianchi type I simplified to a diagonal metric. This is appropriate when one is considering the most likely forms of the stress energy of relevance to the large-scale universe: dust and perfect fluids (see section \ref{bianchi:diagonal}), but is less relevant for our purposes as discussed below. 

\subsection{Kasner solution}\label{kasner}
The unique vacuum Bianchi type I solution is known as the Kasner metric \cite{Kasner}\index{Bianchi type I!Kasner solution},
\begin{equation}\label{kasnerdef}
\d s^2 = -\d t^2 +\sum_i t^{2p_i}\d x^i,
\end{equation}
with constraints
\begin{equation}
\sum_i p_i = 1; \qquad \sum_i p_i^2 = 1.
\end{equation}
Choosing $p_1 \leq p_2 \leq p_3$, the exponents have the ranges 
\begin{align}
-\frac{1}{3} \leq p_1 \leq 0; \qquad
0 \leq p_2 \leq \frac{2}{3}; \qquad
\frac{2}{3} \leq p_3 \leq 1.
\end{align}
Furthermore they can be expressed in terms of the single parameter $u$:
\begin{align}
p_1 = \frac{-u}{1+u+u^2}; \qquad
p_2=\frac{1+u}{1+u+u^2}; \qquad
p_3=\frac{u+u^2}{1+u+u^2}.
\end{align}
This obviously includes the flat three space $\left(p_1, p_2, p_3\right)=\left(0, 0, 1\right)$. Other than this solution, one of the exponents must always be negative; as $t \to 0$ the universe shrinks along two of the primary axes whilst growing along the other. Generally this must also happen at different rates: the only other case where two exponents are the same is $\left(p_1, p_2, p_3\right)=\left(-\frac{1}{3}, \frac{2}{3}, \frac{2}{3} \right)$. 

The Kasner metric has been studied as one of the simplest possible cosmologies, and also in understanding aspects of the approach to singularity in more general cosmologies. 


\section{Ultra-local limits}\index{ultra-local limit}

Simply put, the ultra-local limit is the first step in a possible strong coupling or high-temperature expansion for some field theory. The standard, most well-developed expansion for field theories is a weak-coupling expansion. There is, in addition, a strong-coupling expansion, which has the ultra-local limit as its zeroth term.

This strong coupling expansion has been well examined in the case of quantum field theory (see \cite{Bender, Francisco:quasi2} and chapter 10 of \cite{Klauder1}) and has also been investigated as a starting point for attempts at quantising gravity \cite{Francisco, Francisco:quasi1, Isham, Pilati, Tsacoyeanes}.

Essentially this strong-coupling expansion is done by perturbatively adding small gradient terms to an independent evolution of separate spatial points. It is thus seen that this can be viewed, from a statistical mechanics perspective, as a high-temperature limit.

To make this mathematically precise, in the case of gravity, take the Langragian for the ADM decomposition,\index{Lagrangian}
\begin{equation}
\mathcal{L} = N\; {^{(3)}}g^{1/2}\left(K_{ij}\,K^{ij} - K^2+{^{(3)}}R\right).
\end{equation}

The ultra-local limit corresponds to setting the potential term, ${{}^{(3)}}R$, to zero. As the extrinsic curvatures\index{extrinsic curvature} have only derivatives with respect to time, all spatial derivatives have dropped out and thus each point has decoupled from every other point, and evolves separately.

We could also explicitly reintroduce units,
\begin{equation}
\mathcal{L} = N\; {^{(3)}}g^{1/2}\left(\frac{16\pi G_N}{c^3}\,K_{ij}\,K^{ij} - \frac{16\pi G_N}{c^3}\,K^2+\frac{c^3}{16\pi G_N}\,{^{(3)}}R\right),
\end{equation}
so we see that one way of implementing this ultra-local limit is by taking the coupling constant, $G_N \to \infty$, making it a strong coupling limit. Alternatively, we could see this as taking the speed of light to zero. This forces lightcones to become vertical, so that no two points can interact.

\section{Approach to singularity}

Bianchi cosmologies, and type I cosmologies in particular, have often been of interest in examining the approach to a general singularity. The general idea is that close to the singularity the zeroth approximation to a general spacetime is a Kasner metric, with spatial curvature terms added on in some simple way. There are two ways this could happen; a smooth approach to the singularity (velocity-dominated) or chaotic (Mixmaster behaviour). In general, close to \emph{classical} singularities, it is believed that either \cite{Damour}:
\begin{itemize}

\item The cosmology is Kasner-like.

\item The cosmology is oscillatory, ``generalized Mixmaster'' behaviour.

\item Something else.

\end{itemize}
With many believing that generically the behaviour is Mixmaster. There has been a significant amount of work into discovering when which behaviour happens and if and when the ``something else'' might occur.


\subsection{Velocity dominated limits}
\index{velocity dominated limit}\index{cosmology}

A velocity dominated cosmology, (which is also referred to as having a velocity dominated singularity, or being an asymptotically velocity-term dominated (AVTD) cosmology), is a cosmology that can be approximated on a coordinate patch by a simpler, ultra-local cosmology near the initial singularity \cite{Eardley, Liang2, Liang3}. Denote an ultra-local metric by $\bar g_{ab}$ and extrinsic curvature by $\bar K_{ab}$. A cosmology, $g_{ab}$ with extrinsic curvature $K_{ab}$ is velocity dominated if, when put into synchronous coordinates, in some precise way (for details see \cite{Eardley, Liang3}), as $t \to 0$,
\begin{equation}
g_{ab} \to \bar{g}_{ab}; \qquad K_{ab} \to \bar{K}_{ab},
\end{equation}

This implies that close to the singularity the potential terms can be ignored, so the evolution can be approximated by the kinetic terms \cite{Francisco}. 

Note that this approximation changes the form of matter under consideration, discarding terms in the stress-energy to permit one to approximate by the ultra-local metric. 

Of particular interest is the case when the cosmology is approximated by a Bianchi type I metric near the singularity. These cosmologies are sometimes referred to as Kasner-like, or as exhibiting Kasner-like behaviour \cite{Damour}. A large number of spacetimes are, or are conjectured to be, Kasner-like \cite{Berger}.

A possible source of confusion is that sometimes Mixmaster behaviour (see below) is referred to as being velocity dominated \cite{Rendallmix}, though many researchers regard these as competing behaviours \cite{Damour, Weaver}. 


\subsection{BKL singularities and Mixmaster dynamics}\index{Mixmaster dynamics|(}\index{Bianchi type IX}\index{BKL singularity|(}

A cosmological concept related to these ideas has emerged from examinations of the vacuum Bianchi type IX spacetime (type VIII is often also included in these calculations as the behaviour is very similar), examined independently by Misner \cite{Misner:1969} and Belinsky, Khalatnikov and Lifshitz \cite{BKL}. 

For vacuum Mixmaster, we use $\lambda = \nu = \mu =1$ in equations (\ref{bianchiR1})--(\ref{bianchiR3})\index{vacuum equations}. That is:
\begin{align}
2\ddot \alpha &= \left(b^2 - c^2\right)^2 - a^4;\label{mixaddot} \\
2\ddot \beta &= \left(a^2 - c^2\right)^2 - b^4; \\
2\ddot \gamma &= \left(a^2 - b^2\right)^2 - c^4;
\end{align}
\begin{equation}\label{mixconstraint}
\dot \alpha \dot \beta + \dot \alpha \dot \gamma + \dot \beta \dot \gamma = \frac{1}{4}\left(a^4+b^4 +c^4- 2 a^2 b^2- 2 a^2 c^2- 2b^2 c^2\right).
\end{equation}
If there is some point in time when the righthand side of equations (\ref{mixaddot})--(\ref{mixconstraint}) could be neglected, Kasner behaviour would occur, so that
\begin{equation}
a \propto t^{p_1}; \qquad b \propto t^{p_2}; \qquad c \propto t^{p_3}.
\end{equation}
However, the righthand side will act as a perturbation, so the spacetime would not stay in this Kasner regime forever. We know from an examination of the Kasner spacetimes (section (\ref{kasner})) that in this case as $t \to 0$ two of the terms would decrease (contracting to the big bang) and the other would increase, so the effect of the perturbation will act through the term that increases. Arbitrarily pick this to be $a$. Now,
\begin{equation}
\ddot \alpha = -\frac{1}{2} e^{4\alpha}; \qquad \ddot \beta = \ddot \gamma = \frac{1}{2} e^{4\alpha},
\end{equation}
which can be explicitly integrated:
\begin{align}
a^2 & = \frac{2 |p_1|\Omega}{\cosh(2 |p_1|\Omega \tau)}; \\
b^2 & = b_0^2 \exp(2\Omega\left(p_2- |p_1| \tau \right))\cosh(2 |p_1|\Omega); \\
c^2 & = c_0^2 \exp(2\Omega\left(p_3- |p_1| \tau \right))\cosh(2 |p_1|\Omega).
\end{align}
If we are only interested in the behaviour close to the singularity, then we can approximate these to
\begin{align}
a & \propto \exp(-\Omega \,p_1 \tau); \\
b & \propto \exp(\Omega (p_2+2p_1) \tau); \\
c & \propto \exp(\Omega (p_3+2p_1) \tau);
\end{align}
with
\begin{equation}
t \propto \exp(\Omega (1+2p_1) \tau).
\end{equation}
So it is possible to say that
\begin{equation}
a \propto t^{p'_l}; \qquad b \propto t^{p'_m}; \qquad c \propto t^{p'_n},
\end{equation}
for all times (close to the singularity) with an explicit mapping between the primed and unprimed exponents. For details see \cite{BKL, Montani}. Thus we see that perturbing a Kasner mode leads to another, different Kasner mode, allowing the simplification of a Mixmaster universe into a series of these Kasner modes. This mapping is such that the exponents act like Kasner exponents for significant lengths of time, with sudden changes to new exponents. Furthermore there are generically an infinite number of these oscillations as $t \to 0$. 

It was conjectured \cite{BKL} that this behaviour describes the approach to singularity not just for vacuum Bianchi type IX, but for generic homogeneous and inhomogenous spacetimes containing most relevant forms of matter. Much investigation has gone into trying to determine how true this conjecture is. Essentially it is argued that the Mixmaster dynamics of Bianchi IX spacetimes characterise a generic classical big bang. This is the idea of the BKL singularity. 

In the case of inhomogeneous cosmologies, it is believed that generically the approach to singularity will demonstrate Mixmaster behaviour in the sense that different spatial points will approach different Mixmaster solutions \cite{Weaver}. However, few investigations have looked at whether Mixmaster behaviour occurs in inhomogenous cosmologies. 

It is important to keep in mind that these arguments were developed heuristically, and as such a number of opposing ideas have developed over the many years this issue has been investigated. A collection of useful recent articles regarding BKL singularities and mixmaster dynamics are \cite{Damour, Heinzlemix, Montani, Weaver}.


\subsection{Kinetic approximations}

We see from the above that it has often been desirable to approximate a general spacetime by just the kinetic terms in some appropriate way. Under certain conditions a generic spacetime can be approximated by a simpler one, containing only kinetic terms, after which spatial gradient terms could be added on perturbatively. In the case of velocity dominated limits, the potential terms can be thrown away close to the singularity, while with Mixmaster behaviour the potential terms are important only over brief intervals where they act to change between Kasner modes. \index{Mixmaster dynamics|)}\index{BKL singularity|)}

\section{Discussion}

We see that several limits have been developed which involve some form of ultra-locality, in which Bianchi type I spacetimes have frequently been of central importance. Furthermore taking a limit along a null geodesic has generated significant interest. In the next chapter we combine these two developments to construct an ultra-local limit along a timelike geodesic.

\chapter[Any spacetime has a Bianchi type I cosmology as a limit]{Any spacetime\\has a Bianchi type I cosmology\\as a limit}\label{bianchi}

In this chapter we will construct a particular ultra-local limit along an arbitrary timelike geodesic, research that was presented in \cite{Cropp2}. The limit we describe is closely related to the Penrose limits, and we shall explore the properties of this limit and its resulting spacetime. 

\section{Construction of the limit}

To define the ultra-local\index{ultra-local limit} limit pick an arbitrary timelike geodesic, $\gamma$, and, in some timelike tube surrounding the chosen curve, choose a specific set of adapted coordinates\index{adapted coordinates} coordinates $(t,x^i)$ such that:\index{geodesic!timelike}
\begin{itemize}
\item The curve in question lies at $x^i = 0$.
\item At $x^i = 0$ the coordinate $t$ is just proper time along the curve.
\item The coordinate system is synchronous.\index{synchronous coordinates}
\end{itemize}
This coordinate choice is enough to set
\begin{equation}
\d s^2 = - c^2\, \d t^2 + g_{ij}(t,x) \; \d x^i \d x^j,
\end{equation}
where we have explicitly introduced the speed of light. Note that this coordinate system is only required \emph{locally}, not globally, and hence this choice does not impose any restriction on the type of spacetime being considered. Similarly to section (\ref{penrose:construction}) this now corresponds to embedding the geodesic into a congruence of twist-free timelike geodesics. 

\vspace{20pt}

\noindent
Now implement the ultra-local limit in stages:
\begin{enumerate}
\item 
Make the coordinate transformation $x^i \to \epsilon x^i$. Then
 \begin{equation}
 \d s^2 \to - c^2 \d t^2 + \epsilon^2 g_{ij}(t,\epsilon x) \; \d x^i \d x^j.
 \end{equation}
\item
Change the speed of light $c \to \epsilon c$. So the net effect is 
\begin{equation}
\d s^2 \to - \epsilon^2 c^2 \d t^2 + \epsilon^2 g_{ij}(t,\epsilon x) \; \d x^i \d x^j.
\end{equation}
\item
Perform a conformal transformation $ \d s^2 \to \epsilon^{-2} \d s^2$. So overall
\begin{equation}
\d s^2 \to - c^2 \d t^2 + g_{ij}(t,\epsilon x) \; \d x^i \d x^j.
\end{equation}
\item 
Take the limit $\epsilon\to 0$. Then the net effect is 
\begin{equation}
\d s^2 \to - c^2 \d t^2 + g_{ij}(t) \; \d x^i \d x^j.
\end{equation}
\end{enumerate}
\index{Bianchi type I}

This definition is clearly ``ultra-local''\index{ultra-local limit} --- one has effectively set $c\to 0$ so that nothing propagates, such that each point in ``space'' becomes disconnected from every other point and evolves separately; and each individual point in space is then ``blown up'' by the conformal transformation to yield an independent universe. Essentially we have taken this limit of the spacetime to be given by the geodesic itself, isolated from the influence of the rest of the spacetime. 

\vspace{20pt}

\noindent
An alternative (but completely equivalent) picture of our limit can be constructed by again starting from
\begin{equation}
\d s^2 = - c^2 \d t^2 + g_{ij}(t,x) \; \d x^i \d x^j.
\end{equation}
Consider the decomposition of $g_{ab}$, as in section (\ref{raychaud}), into a projection metric, $h_{ab}$, and timelike vectors $u_a$:
\begin{equation}
g_{ab}(t, \vec x)=h_{ab}(t, \vec x)-u_au_b,
\end{equation}
with
\begin{equation}
h_{ab}= \left[\begin{array}{c|c} 0 & 0 \\ \hline 0 & g_{ij} \end{array} \right]; \qquad u_a=(c,0,0,0).
\end{equation}
Now consider the three dimensional slice
\begin{equation}
\d q^2 = h_{ab}(t,\vec x)\d x^a \d x^b,
\end{equation}
and again implement the limit in stages 

\begin{enumerate}

\item Take the coordinate transformation
\begin{equation}
t \to t; \qquad x^i \to \epsilon x^i,
\end{equation}
resulting in
\begin{equation}
\d q^2 = \epsilon^2 h_{ab}(t,\epsilon \vec x)\d x^a \d x^b.
\end{equation}

\item Perform a conformal transformation
\begin{equation}
\d q^2_\epsilon = \frac{\d q^2}{\epsilon^2} = h_{ab}(t,\epsilon \vec x)\d x^a \d x^b.
\end{equation}

\item Now considering
\begin{equation}
\d s^2_\epsilon =-c^2\d t^2 + \d q^2_\epsilon,
\end{equation}
take the limit $\epsilon \to 0$
\begin{equation}
\d s^2_\epsilon = -c^2\d t^2 + h_{ab}(t,0)\d x^a \d x^b;
\end{equation}
or
\begin{equation}
\d s^2_\epsilon = -c^2\d t^2 +g_{ij}(t,0)\d x^i \d x^j. 
\end{equation} 
\end{enumerate}
From this point of view we have decomposed the spacetime and then taken the flat space limit as in section (\ref{flat}) on the \emph{three-dimensional slice only}. \index{flat space limit}In this coordinate system, we are left with the value of the metric along the timelike geodesic. 

\vspace{20pt}
\noindent
Note that the output from this ultra-local limit is exactly the definition of a general Bianchi type I spacetime --- the general definition before you make any assumptions about diagonalisability of the spatial metric $g_{ij}(t)$. Also note that if one starts with a congruence of timelike geodesics in the original spacetime, then the output of this ultra-local limit is a collection of general Bianchi type I spacetimes, one Bianchi spacetime being attached to each timelike geodesic. \index{Bianchi type I}

\section{Features and relation to other limits}

Clearly this limit closely parallels the Penrose limit construction, in fact one can argue that it is as close as one can possibly get to a ``timelike Penrose limit'' --- though one should certainly not expect all of the features of the usual ``lightlike Penrose limit'' to survive in this timelike case. \index{Penrose limit}

Note carefully that this ultra-local limit does not share the hereditary properties discussed in section (\ref{penrose:hereditary})\index{hereditary properties} of the ordinary Penrose limit. The information about the original spacetime retained or discarded in the two limits will be different.\index{Penrose limit} In particular, note the ultra-local limit of a Ricci flat spacetime is not necessarily Ricci flat --- from the ADM\index{ADM decomposition} viewpoint (described in section \ref{ADM}) this is with hindsight obvious since during the limiting process the extrinsic curvatures and the intrinsic (spatial) curvatures scale in a different manner. 

\vspace{20pt}
\noindent
In terms of the physical usefulness of this limit, it is best looked at in the context of two other well-known limits:

\begin{itemize}

\item If we are interested in approximating a spacetime \emph{at a point} the appropriate construction would be to choose Riemann normal coordinates and then approximate the spacetime by the tangent space at that point, as described in section (\ref{flat}).\index{flat space limit}

\item If one is interested in approximating a spacetime \emph{along a null geodesic} then the appropriate construction is to use Penrose's adapted coordinates and take the Penrose limit along that null geodesic as described in section (\ref{penroselim}).\index{Penrose limit}

\item Likewise, if one wishes to approximate a spacetime \emph{along a timelike geodesic}, the appropriate choice is to use synchronous coordinates and then take the limit described above along the timelike geodesic as described in the process above.

\end{itemize}

\section{General Bianchi type I spacetime}
Now that the speed of light $c$ has done its job, let us again adopt units where $c=1$, so that we are considering
\begin{equation}
\d s^2 = -\d t^2 + g_{ij}(t) \; \d x^i \d x^j.
\end{equation}
Geometrically this is a rather simple spacetime. Define the extrinsic curvature \index{extrinsic curvature}
\begin{equation}\label{K}
K_{ij} = -\frac{1}{2} \frac{\d g_{ij}}{\d t},
\end{equation}
then in particular
\begin{equation}\label{tracek}
\tr(K) = - \frac{1}{\sqrt{g_3}} \; \frac{\d \sqrt{g_3}}{\d t},
\end{equation}
where the notion of ``trace'' implicitly involves appropriate factors of the 3-metric and its inverse. We can specialise the usual results for the ADM decomposition  (for an explicit computation, see (and appropriately modify) the discussion of the ADM decomposition in Misner, Thorne, and Wheeler \cite{MTW}), to determine the useful geometric tensors for the Bianchi type I spacetime. 

\subsection{Riemann tensor}\index{Riemann tensor}

The key results for the Riemann tensor are 
\begin{equation}
R_{ijkl} = K_{ik} K_{jl}- K_{il} K_{jk};  
\end{equation}
and 
\begin{equation} 
R_{0i0j} = \frac{\d K_{ij}}{\d t} + K_{im} K^m{}_j = \frac{\d K_{ij}}{\d t} + (K^2)_{ij};
\end{equation}
with all other components zero.

\subsection{Ricci tensor and scalar}\index{Ricci tensor}

A brief computation yields
\begin{eqnarray}
R_{00} 
&=& \frac{\d \tr(K)}{\d t } - \tr(K^2),
\end{eqnarray}
while
\begin{eqnarray}
R_{ij} 
& =& - \frac{\d K_{ij}}{\d t} - 2 K_{im} K^m{}_j + \tr(K) K_{ij}.
\end{eqnarray}
Therefore
\begin{eqnarray}
R 
&=& -2\frac{\d \tr(K)}{\d t}+  [\tr(K)]^2 + \tr(K^2).
\end{eqnarray}

\subsection{Einstein tensor}\index{Einstein tensor}

The Einstein tensor can also be evaluated from the above, or can be read off from Misner, Thorne, and Wheeler \cite{MTW} (see (21.162 a,b,c) p 552). The time-time component is simple
\begin{eqnarray}
G_{00} 
= 
\frac{1}{2} \left\{ [\tr(K)]^2 - \tr(K^2) \right\}.
\end{eqnarray}
In contrast the space-space components are relatively messy
\begin{align}\label{spacespace}
G_{ij} = &- \frac{\d}{\d t} \left[ K_{ij} - g_{ij} \tr(K) \right] + 3 \tr(K) K_{ij} - 2 (K^2)_{ij} \nonumber \\
&- \frac{1}{2} g_{ij} \left\{  [\tr(K)]^2 + \tr(K^2) \right\}.
\end{align}\\
\section{Simplifying the general Bianchi type I spacetime}\index{Bianchi type I}

These calculations for the relevant tensors of the general Bianchi type I spacetime are reasonably simple but there are several other tricks that may be employed to put them into forms useful for further exploration of these spacetimes.\\

\subsection{Mixed components}

It is often advantageous to work with \emph{mixed} components such as $R^a{}_b$. Working from the above, (or suitably adapting results from the key article on BKL spacetimes \cite{BKL}), one has:\index{Ricci tensor}
\begin{equation}
R^0{}_0 = - \frac{\d \tr(K)}{\d t } + \tr(K^2);
\end{equation}
\begin{equation}
R^i{}_0 = 0; \qquad R^0{}_i = 0;
\end{equation}
\begin{equation}\label{rijmixed}
R^i{}_j = -\frac{1}{\sqrt{g_3}} \frac{\d}{\d t} \left[ \sqrt{g_3} \; K^i{}_j \right].
\end{equation}
Also note from equation (\ref{rijmixed}) and using equation (\ref{tracek}) that (summing only over the space indices)
\begin{equation}
R^i{}_i =\frac {1}{\sqrt{g_3}} \frac{\d^2}{\d t^2} \sqrt{g_3}. 
\label{E:xxx}
\end{equation}
For the Einstein tensor, the space-space component (equation (\ref{spacespace})) simplifies somewhat, becoming\index{Einstein tensor}
\begin{equation}
G^i{}_j = -\frac{1}{\sqrt{g_3}} \frac{\d}{\d t} \left[ \sqrt{g_3} \; K^i{}_j \right] 
+ \delta^i{}_j \left\{ \frac{\d \tr(K)}{\d t}  - \frac{[\tr(K)]^2}{2} - \frac{\tr(K^2)}{2} \right\}.
\end{equation}

\subsection{Unimodular decomposition}\index{unimodular decomposition}
We will now attempt a decomposition similar to that in section (\ref{rosen:decompose});
\begin{equation}
g_{ij}(t) = a(t)^2 \; \hat g_{ij}(t),\index{scale factor}
\end{equation}
where $\det(\hat g_{ij} ) = 1$. Here $a(t)$ can be viewed as an overall scale factor (similar to that occurring in FLRW cosmologies), while $\hat g_{ij}$ describes the ``shape'' of space.
Then
\begin{equation}
\frac{\d g_{ij}}{\d t} = 2 a \dot a \; \hat g_{ij} + a^2\; \frac{\d{\hat g_{ij}}}{\d t},
\end{equation}
and so from equation (\ref{K}),
\begin{equation}
K_{ij} = - a \dot a \; \hat g_{ij} + a^2 \; \hat K_{ij},
\end{equation}
with $\hat g^{ij} \,\hat K_{ij} = 0$ because of the unit determinant condition. 
Now define 
\begin{equation}
\hat K^i{}_j = \hat g^{ik} \, \hat K_{kj}
\end{equation}
so that
\begin{equation}
K^i{}_{j} = - \frac {\dot a}{ a} \; \delta^i{}_{j} + \hat K^i{}_{j}.
\end{equation}
Thus
\begin{equation}
\tr(K) = - 3\; \frac{\dot a}{a};
\end{equation}
\begin{equation}
\tr(K^2) = 3 \; {\dot a^2}{ a^2} + \tr( \hat K^2 ).
\end{equation}
Then\index{Ricci tensor}
\begin{equation}
R_{00} = 3 \frac{\ddot a}{ a } + \tr(\hat K^2);
\end{equation}
and 
\begin{equation}
R_{ij} = a^2 \left\{ -\frac {\d \hat K_{ij}}{ \d t} - 2 \hat K_i{}^m \hat K_{mj} + \tr(\hat K) \hat K_{ij} \right\} + ( a \ddot a + 2 \dot a^2 ) \hat g_{ij},
\end{equation}
whilst the Einstein tensor is given by\index{Einstein tensor}
\begin{equation}
G_{ij} = a^2 \hat G_{ij} + a \dot a \hat K_{ij}  -(2a\ddot a + \dot a^2) \hat g_{ij}.
\end{equation}

We find these do not simplify as well as the expression in equation (\ref{rosen:d}), as we no longer have a trace over all indices. However it is possible to make a slight improvement by combining this decomposition with the previous use of mixed components. Then
\begin{equation}
R^0{}_0 = 3 \frac{\ddot a}{ a } + \tr(\hat K^2);
\end{equation}
and
\begin{equation}
R^i{}_j = -\frac {\d}{\d t} \left[\hat K^i{}_j \right] - 3\frac {\dot a}{ a} \hat K^i{}_j 
+ \left\{\frac{\ddot a}{ a} + 2\frac {\dot a^2}{ a^2} \right\} \delta^i{}_j.
\end{equation}
Summing over the space components only we have (in agreement with the previous subsection, see equation (\ref{E:xxx}))
\begin{equation}
R^i{}_i =
3 \left\{\frac{\ddot a}{ a} + 2 \frac{\dot a^2}{ a^2} \right\} = \frac{1}{a^3} \frac{\d^2(a^3)}{\d t^2}.
\end{equation}
For the Einstein tensor the interesting piece is
\begin{equation}
G^i{}_j = - \frac{\d}{\d t} \left[\hat K^i{}_j \right] - 3 \frac{\dot a}{a} \hat K^i{}_j 
- \frac{1}{2}\left\{\frac{\ddot a}{a} + 2\frac{\dot a^2}{a^2} \right\} \delta^i{}_j.
\end{equation}
\subsection{Summary}

We have attempted to find a simple way to write down various useful tensors for the general Bianchi type I spacetime. The spacetime curvature of the general Bianchi type I spacetime can be evaluated as simple algebraic combinations of the $3\times3$ matrices $g_{ij}$, $\dot g_{ij}$, and $\ddot g_{ij}$. If we split the geometry into an overall scale factor $a(t)$ and a shape $\hat g_{ij}$, then the curvature can be evaluated as a simple algebraic combination of $a$, $\dot a$, $\ddot a$ and the $3\times3$ matrices $\hat g_{ij}$, $\dot {\hat g}_{ij}$, and $\ddot {\hat g}_{ij}$.

Due both to ease of physical interpretation and simplicity of equations it is often useful to work with mixed components, and to split the metric into a scale factor and a unit determinant piece.

\section{Stress-energy in the ultra-local limit}\index{stress-energy tensor}

What happens to the stress-energy in the ultra-local limit we are interested in? Based on physical intuition one expects that for the energy-momentum flux $T_{0i} \to 0$, as the limit means that each point is disconnected from all others, evolving independently, but what about $T_{00}$ and $T_{ij}$? 
Remember that for the light-like Penrose limit the stress-energy also changes as $T_{ab}\to T_{uu}$\index{Penrose limit}, or alternatively $T_{ab} \to k_ak_b$ with $k$ some null vector, with all other components vanishing \cite{blaulecture2, Penrose1, Penrose2, steele}.\index{ultra-local limit}

It is convenient to develop the ultra-local limit in a somewhat different form: 

\begin{enumerate}
\item Make the replacement
\begin{equation}
g_{ab}(t,x) \to g^\epsilon_{ab}(t,x) = g_{ab}(t, \epsilon x).
\end{equation}
\item Furthermore, for any generic field $\Psi(t,x)$ make the replacement
\begin{equation}
\Psi(t,x) \to \Psi_\epsilon(t,x) = \Psi(t, \epsilon x).
\end{equation}
\item Then consider the limit as $\epsilon\to0$.
\end{enumerate}

We consider a few specific cases. Most of the forms of stress-energy described below were considered in section (\ref{matter}).

\subsection{Perfect fluid}\index{perfect fluid}
Consider a generic perfect fluid in a generic spacetime,
\begin{equation}
T^{ab} = (\rho+p) V^a V^b + p g^{ab},
\end{equation}
and ask what the ultra-local limit might be? Working from the geometrical side we have $G^{0i}\to 0$ in the naturally adapted coordinate system. Thus via the Einstein equations it follows that we must also have $T^{0i} \to 0 $, which in turn implies $V^i \to 0$. Therefore 
\begin{equation}
T_{00} \to \rho;
\qquad
T_{0i} \to 0;
\qquad
T_{ij} \to p \; g_{ij}.
\end{equation}
That is, the generic perfect fluid reduces in the ultra-local limit to a co-moving perfect fluid.

\subsection{Scalar field} \index{scalar field}

Consider a scalar field $\phi(t,x)$, which has stress-energy tensor
\begin{equation}
T_{ab} = \phi_{,a} \phi_{,b} - \frac{1}{2}g_{ab} \left\{ (\nabla\phi)^2 + V(\phi) \right\}.
\end{equation}
Now take the ultra-local limit. Note 
\begin{align}
T_{00}^{\epsilon} =&\, \frac{1}{2}\left\{ \dot \phi_\epsilon^2 + g^{ij}_\epsilon \partial_i \phi_\epsilon \partial_j \phi_\epsilon + V\right\} \nonumber \\
=& \,\frac{1}{2}\left\{ \dot \phi^2 + \epsilon^2 g^{ij} \partial_i \phi \partial_j \phi +V \right\} \nonumber\\
\to &\, \frac{1}{2} \left\{ \dot \phi^2(t,0) + V \right\},
\end{align}
whence ultimately
\begin{equation}
T_{00}^{\epsilon} \to \frac{1}{2} \left\{ \dot \phi^2(t,0) + V \right\}.
\end{equation}
Similarly
\begin{equation}
T_{0i}^{\epsilon} = \frac{1}{2}\left\{ \dot \phi_\epsilon \partial_i \phi_\epsilon\right\} \to \frac{1}{2}\left\{ \epsilon \dot \phi \partial_i \phi \right\} \to 0.
\end{equation}
Furthermore
\begin{align}
T_{ij}^{\epsilon} &= \partial_i \phi_\epsilon \partial_j \phi_\epsilon - \frac{1}{2} g_{ij}^{\epsilon} \left\{ (\nabla\phi)^2 + V \right\} \nonumber\\
&\to \epsilon^2 \partial_i \phi \partial_j \phi - \frac{1}{2}\left\{ (-\dot\phi^2 + \epsilon^2 (\partial\phi)^2 ) + V \right\},
\end{align}
whence ultimately
\begin{equation}
T_{ij}^{\epsilon} \to \left\{ \dot \phi^2(t,0) - V \right\} g_{ij}.
\end{equation}
The key point is that the structure of the stress-energy tensor guarantees that in the ultra-local limit 
\begin{equation}
T_{00} \to \rho;
\qquad
T_{0i} \to 0;
\qquad
T_{ij} \to p \; g_{ij}.
\end{equation}
Similarly to the limit of the generic perfect fluid, the scalar field situation results in a co-moving perfect fluid. Such simple behaviour is common, but by no means universal.

\subsection{Electromagnetic field} \index{electromagnetic field}

For the electromagnetic field the most natural form of the ultra-local limit is to consider the vector potential of equation (\ref{vecpot}) and take\index{vector potential}
\begin{equation}
A^a(t,x) \to A_\epsilon^a(t,x) = \left( \phi(t,\epsilon x); \; \vec A(t,\epsilon x) \right),
\end{equation}
in which case 
\begin{equation}
\vec E \to - \dot {\vec A}(t, 0); \qquad \vec B \to 0.
\end{equation}
So for the stress-energy tensor,
\begin{equation}
T_{00} \to \frac{1}{2} \left\{ g^{ij} \dot A_i \dot A_j\right\};
\end{equation}
\begin{equation}
T_{0i} \to 0;
\end{equation}
\begin{equation}
T_{ij} \to E_i E_j -  \frac{1}{2} \left\{ g^{kl} E_k E_l\right\} g_{ij}.
\end{equation}
In terms of the electric field $E$, and the naturally defined inner product $g(E,E)$, it is best to summarise this as:
\begin{equation}
T_{00} \to \frac {1}{2} \; g(E,E);
\end{equation}
\begin{equation}
T_{0i} \to 0;
\end{equation}
\begin{equation}
T_{ij} \to E_i E_j- \frac {1}{2} \; g(E,E) g_{ij}.
\end{equation}
Note that in general the $3\times3$ matrix of space-space components of the stress-energy will not be diagonal. Thus this gives an important case where many of the cosmological calculations cannot be blindly followed. 

Note further that in reference \cite{Ryan} a somewhat different limit is taken, because those authors are interested in different physics. This is done because if one is interpreting the Bianchi type I spacetime as a cosmology, as an approximation to the large-scale structure of our own physical universe, then for physical reasons one might wish to allow $B\neq0$, while expecting that $ E\to0$, as magnetic fields may be important in cosmology but electric fields almost certainly are not. Nevertheless, even in this very different context, one key point remains: The $ 3\times3$ matrix of space-space components of the stress-energy need not and generally will not be diagonal.

\subsection{Anisotropic fluid}\index{anisotropic fluid}
Consider a generic anisotropic fluid 
\begin{equation}
T^{ab} = (\rho+p) V^a V^b + p g^{ab} + \Theta^{ab}.
\end{equation}
Here $\Theta$ is the anisotropic stress tensor, which is always 4-traceless and 4-orthogonal to the 4-velocity $V$. What might the ultra-local limit be? However we choose to define this limit, since we know from the geometrical side that $G^{0i}\to 0$, we must via the Einstein equations have $T^{0i} \to 0 $, so that the stress-energy tensor is (3+1) block diagonal. This in turn implies that we can without loss of generality redefine our variables so that $V^i \to 0$, while $V^0\to 1$, and simultaneously choose $\Theta^{00}\to 0$ along with $\Theta^{0i}\to 0$. Therefore 
\begin{equation}
T_{00} \to \rho;
\qquad
T_{0i} \to 0;
\qquad
T_{ij} \to \pi_{ij} = p g_{ij} + \Theta_{ij}.
\end{equation}
Once again note we do not necessarily have a diagonal metric for more general forms of matter. 

\subsection{Stress-energy conservation}
Consider the covariant conservation law
\index{conservation law}
\begin{equation}
\nabla_a T^{ab}=0 \implies \frac{1}{\sqrt{-g_4}} \partial_a ( \sqrt{-g_4} \; T^{ab}) + \Gamma^b{}_{cd} T^{cd} = 0.
\end{equation}
For Bianchi type I the only nontrivial component is
\begin{equation}
 \frac{1}{\sqrt{g_3}} \partial_t ( \sqrt{g_3} \; T^{00}) + \Gamma^0{}_{cd} T^{cd} = 0,
\end{equation} 
which implies 
\begin{equation}
 \frac{1}{\sqrt{g_3}} \partial_t ( \sqrt{g_3} \rho) + \frac{1}{2}\frac{\d g_{ij}}{\d t} \pi^{ij} = 0,
\end{equation} 
that is
\begin{equation}
 \frac{1}{\sqrt{g_3}} \partial_t ( \sqrt{g_3} \rho) = -\frac{1}{2}\frac{\d g_{ij}}{\d t} \pi^{ij}.
\end{equation} 
To see the connection with the well understood FLRW spacetime (section \ref{penrose:FLRW}) note $g_{ij} = a^2 \; \hat g_{ij}$ and then
\begin{equation}
\frac{1}{ a^3} \; \partial_t ( a^3 \rho) = 
 -\frac{1}{2} \left(2 a\dot a \,\hat g_{ij} - a^2 \,\frac{\d\hat g_{ij}}{\d t} \right) \pi^{ij},
\end{equation} 
which we can write as
\begin{equation}
\partial_t ( a^3 \rho) =  -a^2 \dot a g_{ij} \pi^{ij} -  \frac{1}{2}a^5 \frac{\d\hat g_{ij}}{\d t} \pi^{ij}.
\end{equation} 
That is, (defining $\pi_{ij} = a^2 \, \hat\pi_{ij}$, so that $\pi^i{}_j = \hat \pi^i{}_j$ and $\pi^{ij} = a^{-2} \, \hat\pi^{ij}$),
\begin{equation}
\frac{\d( \rho \, a^3)}{ \d t} =   - 3 a^2 \dot a p -  \frac{1}{2}\; a^3 \; \frac{\d\hat g_{ij}}{\d t} \; \{ \hat \pi^{ij} - p \hat g^{ij}\}.
\end{equation}
Finally, (defining $\Theta_{ij} = a^2 \, \hat\Theta_{ij}$, so that $\Theta^i{}_j = \hat \Theta^i{}_j$ and $\Theta^{ij} = a^{-2} \, \hat\Theta^{ij}$),
\begin{equation}
\frac{\d( \rho \, a^3)}{ \d t} =  - p\frac{\d (a^3)}{\d t}  -  \frac{1}{2}\; a^3 \; \frac{\d\hat g_{ij}}{\d t} \; \hat \Theta^{ij}.
\end{equation}
The first 2 terms are the familiar FLRW conservation law as given in equation (\ref{penrose:d}). The last term involves the trace-free anisotropic part of the (spatial) stress tensor, together with the trace-free contribution from $\d\hat g_{ij}/\d t$. We can also write this as
\begin{equation}
\frac{\d\rho}{\d t} =  - 3(\rho+p) \frac{\d a}{\d t} + \hat K_{ij}\; \hat \Theta^{ij}.
\end{equation}
So the anisotropy modifies the conservation equation, splitting essentially into parts concerning the change in density due to the change in ``size'' and that due to a traceless contribution due only to the change in shape of the space. 

\subsection{Summary} 
In general the message is that in the ultra-local limit the stress-energy satisfies
\begin{equation}
T_{00} \to \rho;
\qquad
T_{0i} \to 0;
\qquad
T_{ij} \to \pi_{ij};
\end{equation}
and that one cannot say anything more than this without specifying the particular form of the matter content. It seems that looking at stress-energy, we obtain a more complex picture than the lightlike Penrose limit, whilst sharing the fact that both limits erase some terms from the stress-energy. Furthermore in this ultra-local limit there is a natural generalisation of the energy conservation law normally applied to FLRW spacetimes, with an extra term coming from the interplay between anisotropies in the stress tensor and changes in the shape (not volume) of the spatial slices.

\section{Einstein equations}\index{Einstein equations}

In the ultra-local limit the Einstein equations reduce to three significant pieces of information --- arising from the the time-time component, the spatial trace, and the anisotropic part of the spatial trace. As in section (\ref{flrwee}), it is most convenient to rewrite the Einstein equations as 
\begin{equation}
R_{ab} = 8\pi G_N \{ T_{ab} - \frac{1}{2}T g_{ab}\},
\end{equation}
in which case
\begin{equation}\label{eineq1}
R_{00} = 4\pi (\rho+3p);
\end{equation}
\begin{equation}\label{eineq2}
R^i{}_i = 12\pi(\rho-p);
\end{equation}
\begin{equation}\label{eineq3}
R_{ij} - \frac{1}{3}R^k{}_k g_{ij} =8\pi \, \Theta_{ij}. 
\end{equation}
The first of these (equation (\ref{eineq1})) implies
\begin{equation}
\frac{\ddot a}{a} = - \frac{4\pi}{3} (\rho+3p) - \frac{1}{3} \tr( \hat K^2 ).
\end{equation}
That is, as compared to FLRW (cf equation \ref{penrose:j})\index{FLRW metric} spacetimes, changes in the shape of the spatial slices can now contribute to deceleration.
The second equation (\ref{eineq2}) yields
\begin{equation}
\frac{\ddot a}{a} + 2 \,\frac{\dot a^2}{a^2} = 4\pi(\rho-p).
\end{equation}
So this combination of terms is what one might expect anyway for any spatially flat FLRW cosmology (cf. equation \ref{penrose:k}). 
Finally we obtain an additional relation from equation (\ref{eineq3}), 
\begin{equation}
- \frac{\d}{\d t} \left[\hat K^i{}_j \right] - 3 \, \frac{\dot a}{a} \, \hat K^i{}_j = 8\pi \, \Theta^i{}_j. 
\end{equation}
Alternatively,
\begin{equation}
- \frac{\d}{\d t} \left[a^3\,\hat K^i{}_j \right] = 8\pi \,a^3 \, \Theta^i{}_j. 
\end{equation}
That is, evolution of the \emph{shape} of the spatial slices is driven by anisotropies in the stress tensor. This is the simplest and most straightforward decoupling of the Einstein equations we have managed to find. 

\section{Conditions for a [block]-diagonal metric}\label{bianchi:diagonal}\index{Bianchi type I!diagonal}

A key issue in taking this ultra-local limit (and in Bianchi type I cosmologies generally) is the question of when it is possible (or desirable) to restrict attention to diagonal spatial metrics (or more generally block diagonal metrics). In the cosmological setting it is common to see the full Bianchi type I universe quickly reduced to the diagonal case. Fortunately, within the context of Bianchi type I spacetimes it is possible to prove the following qualitative results: 
\begin{equation}
\hbox{(diagonal stress-energy)} \Leftrightarrow \hbox{(diagonal metric)}. \nonumber
\end{equation}
\begin{equation}
\hbox{(block diagonal stress-energy)} \Leftrightarrow \hbox{(block diagonal metric)}. \nonumber 
\end{equation}

\paragraph{Proof $(\Leftarrow$):} A [block]-\-diagonal metric $g_{ij}(t)$ implies that $\dot g_{ij}(t)$ is [block]-diagonal. This in turn implies that $K_{ij}(t)$ and $\dot K_{ij}(t)$ are [block]-diagonal. Through the calculations for the Ricci tensor, this implies $R_{ij}(t)$ is [block]-diagonal, and this then implies $G_{ij}(t)$ is [block]-\-dia\-gonal, which implies $\pi_{ij}(t)$ is [block]-\-diagonal, implies $\pi^i{}_j(t)$ is [block]-\-diagonal. (Completely equivalently, this implies $\Theta_{ij}(t)$ is [block]-diagonal, which implies $\Theta^i{}_j(t)$ is [block]-diagonal). 

\paragraph{Proof $(\Rightarrow)$:}
At time $t_0$ there is no loss of generality in picking coordinates such that both $[g_0]_{ij} = \delta_{ij}$, and such that $[\dot g_0]_{ij} $ is diagonal. Then without loss of generality there is a coordinate system such that $[K_0]_{ij}$ and $[K_0]^i{}_j$ are diagonal.

Our definition of [block]-diagonal stress-energy will be to assert that in this particular coordinate system $T^i{}_{j}(t)$ is [block]-diagonal, whence $G^i{}_j(t)$ is [block]-diagonal, and so $R^i{}_j(t)$ is [block]-diagonal. 

Note that we have only imposed a restriction on the metric at one particular time. We show in this case the Einstein equations do not lead to evolution away from a diagonal metric. 

Now, using
\begin{equation}
R^i{}_j = -\frac{1}{\sqrt{g_3}} \frac{\d}{\d t} \left[ \sqrt{g_3} \; K^i{}_j \right].
\end{equation}
which we derived earlier (equation (\ref{rijmixed})), as $R^i{}_j$ is [block]-diagonal we rearrange to
\begin{equation}
\frac{\d}{\d t} [ \sqrt{g_3} K^i{}_j ] \propto \{\hbox{[block]-diagonal}(t)\}^i{}_j,
\end{equation}
and so 
\begin{equation}
\sqrt{g_3} K^i{}_j = \sqrt{g_{3,0}} [K_0]^i{}_j + \{\hbox{[block]-diagonal}(t)\}^i{}_j.
\end{equation}
By our initial assumption on the diagonal nature of $[K_0]^i{}_j$ we have
\begin{equation}
K^i{}_j = \{\hbox{[block]-diagonal}(t)\}^i{}_j.
\end{equation}
That is
\begin{equation}
\forall t:  g^{im} \; \dot g_{mj} = \{\hbox{[block]-diagonal}(t)\}^i{}_j,
\end{equation}
but then 
\begin{equation}
\forall t: \dot g_{ij} =  g_{im} \; \{\hbox{[block]-diagonal}(t)\}^m{}_j,
\end{equation}
which can be formally integrated in terms of a time-ordered product
\begin{eqnarray}
[g(t)]_{ij} &=& [g_0]_{im} \exp\left[ \{\hbox{[block]-diagonal}(t)\}\right]^m{}_j
\\
&=& \delta_{im} \exp\left[ \{\hbox{[block]-diagonal}(t)\} \right]^m{}_j,\label{timeordered}
\end{eqnarray}
where in equation (\ref{timeordered}) we are again using coordinate choice that the initial metric is $ \delta_{im}$. Now
\begin{eqnarray}
[g(t)]_{ij} &=& \exp\left[ \{\hbox{[block]-diagonal}(t)\} \right]_{ij}
\\
&=& \{\hbox{[block]-diagonal}(t)\}_{ij},
\end{eqnarray}
implying that $[g(t)]_{ij}$ remains [block]-diagonal for all time. 

\vspace{25pt}
\noindent
To summarise: a \emph{necessary} and \emph{sufficient} condition for a [block]-diagonal metric (in the context of Bianchi type I spacetimes) is that, in the coordinate system where $[g_0]_{ij}=\delta_{ij}$ and $[K_0]_{ij}$ is diagonal, we have 
\begin{equation}
\pi^i{}_j(t) \hbox{ is [block]-diagonal.}
\end{equation}
The metric and stress tensor then decompose into direct sums:
\begin{equation}
g_{ij} = \oplus_A \, [h_A]_{ij}; \qquad \pi_{ij} = \oplus_A \, \{ p_A [h_A]_{ij} \} ; \qquad \pi^i{}_j = \oplus_A\, \{ p_A [\delta_A]^i{}_{j}\}.
\end{equation}


\subsection{Examples} 
For a fully diagonal metric in three space dimensions the three canonical examples of this behaviour are:
\\
\begin{enumerate}
\item 
Perfect fluid:\index{perfect fluid}\index{dust}
\begin{equation}
\pi_{ij} = p \; g_{ij}.
\end{equation}
That is:
\begin{align}
\pi^i{}_j = \left[\begin{array}{ccc}p& 0&0\\ 0 & p & 0\\ 0&0& p \end{array} \right] = p \; \delta^i{}_j.
\end{align}
The special sub-case $p=0$ corresponds to dust, and the even more special sub-case $\rho=p=0$ corresponds to vacuum (Kasner solutions)\index{Bianchi type I!Kasner solution} --- with all of these special cases (perfect fluid, dust, vacuum) being very well studied in a cosmological setting.\index{perfect fluid}\index{dust}

\item 
Electromagnetic field with the electric field $\vec E$ being an eigenvector of $g$:\index{electromagnetic field}
\begin{equation}
\pi_{ij} = C_1 g_{ij} + C_2 V_i V_j; \qquad V^i \propto V_i.
\end{equation}
This is equivalent to:
\begin{align}
\pi_{ij} = \left[\begin{array}{ccc}p_1 \; g_{11}& 0&0\\ 0 & p_2 \; g_{22} & 0 \\ 0 & 0 & p_2 \; g_{33} \end{array} \right]; \qquad 
g_{ij} = \left[\begin{array}{ccc}g_{11}& 0 & 0\\ 0 & g_{22} & 0 \\ 0 & 0 & g_{33} \end{array} \right]. 
\end{align}
That is:
\begin{align}
\pi^i{}_j = \left[\begin{array}{ccc}p_1& 0&0\\ 0 & p_2 & 0\\ 0&0& p_2 \end{array} \right].
\end{align}

\item General diagonal metric:
\begin{equation}
\pi_{ij} = \diag(p_1,p_2,p_3); \qquad g_{ij} = \diag(g_{11},g_{22},g_{33}).
\end{equation}
This is equivalent to:
\begin{align}
\pi_{ij} = \left[\begin{array}{ccc}p_1 g\;_{11}& 0&0\\ 0 & p_2 \;g_{22} & 0\\ 0&0& p_3 \;g_{33} \end{array} \right]; \qquad 
g_{ij} = \left[\begin{array}{ccc}g_{11}& 0&0\\ 0 & g_{22} &0\\0&0&g_{33} \end{array} \right].
\end{align}
That is:
\begin{align}
\pi^i{}_j = \left[\begin{array}{ccc}p_1& 0&0\\ 0 & p_2 & 0\\ 0&0& p_3 \end{array} \right].
\end{align}
\end{enumerate}

There is also a partially diagonalisable case corresponding to one irreducible $2\times2$ block plus a singleton $1\times1$ block, and the fully generic non-diagonalizable case corresponding to the stress tensor being a single irreducible $3\times3$ block.
These are the only relevant cases for 3-dimensions.

\section{Raychaudhuri equation}\index{Raychaudhuri equation}

In a Bianchi type I spacetime we can simplify the standard (timelike) Raychaudhuri equation \cite{HE, Wald} by noticing that we have a natural timelike congruence defined by $u = \partial_t$, while we also have $u = -(\d t)^\sharp$, or more prosaically $u^a = (1,0,0,0)$. For this timelike congruence the vorticity is zero, $\omega=0$, \index{vorticity}as is the acceleration $\frac{\mathsf d u^a}{\mathsf d s} = u^b \nabla_b u^a =0$. 

Therefore in terms of the expansion tensor $\theta_{ab} = \nabla_{(a} u_{b)}$, (not to be confused with the anisotropic stress $\Theta_{ab}$), we have 
\begin{equation}
R_{ab}u^au^b= -\theta_{ab}\theta^{ab}+\theta^2-\frac{\d \theta}{\d t}.
\end{equation}
Separating out the unit determinant metric, $g_{ij} = a^2 \hat g_{ij}$, and using the definitions of shear and expansion given in section (\ref{raychaud}) we find
\begin{equation}
\theta_{ab}= a \dot a \; \hat g_{ab}+a^2\hat K_{ab}; \index{expansion}
\end{equation}
\begin{equation}
\sigma_{ab}=a^2\hat K_{ab}; \index{shear}
\end{equation}
\begin{equation}
\sigma^2=\frac{1}{2}\tr(\hat K^2).
\end{equation}
Furthermore $\displaystyle K=-\frac{3\dot a}{a}$, and hence 
\begin{equation}
\theta = 3\frac{\dot a}{a};
\qquad
\frac{{\d \theta}}{{\d t}} = 3\frac{\ddot a}{a}-3\frac{\dot a^2}{a^2}.
\end{equation}
So the Raychaudhuri equation is given by
\begin{equation}
R_{ab} u^a u^b=-\tr(\hat K^2)-3\,\frac{\ddot a}{a}.
\end{equation}
Compare this to both the result for the FLRW cosmology, previously given in equation (\ref{FLRW:ray}),\index{FLRW metric}
\begin{equation}
R_{ab}u^au^b=-3\, \frac{\ddot a}{a},
\end{equation}
to which the Bianchi type I spacetime reduces when $\hat K_{ab}=0$, and to the result stated by Collins and Ellis \cite{Collins2} for perfect fluid Bianchi cosmologies,
\begin{equation}
\frac{1}{2}(\rho+3p)=-3\frac{\ddot a}{a}-2\sigma^2.
\end{equation}
Attempting to put the Raychaudhuri scalar into a ``nice'' form, the best possible form seems to be
\begin{equation}
R_{ab}u^au^b=\frac{1}{2}\,(\rho+3p)+\tr(\hat K^2),
\end{equation}
or alternatively
\begin{equation}
R_{ab}u^au^b=\frac{1}{4}\,(\rho+3p)-\frac{3}{2}\,\frac{\ddot a}{a}.
\end{equation}\\

\section{Some examples}

We will now look at a few simple examples of this limit.

\subsection{FLRW}
Let us look at the simplest example of this limit that is not completely trivial. Consider the FLRW spacetime in Cartesian coordinates,\index{FLRW metric}
\begin{equation}
\d s^2 = -\d t^2 + a(t)^2\,\left\{ \frac{\d x^2+\d y^2 +\d z^2}{1+ \frac{k\left(x^2 +y^2 +z^2\right)}{4}}\right\}.
\end{equation}
This is already in the desired adapted, synchronous coordinate system.\index{synchronous coordinates}\index{adapted coordinates}
The only effect of the ultra local limit is that
\begin{equation}
1+ \frac{k\left(x^2 +y^2 +z^2\right)}{4} \to 1,
\end{equation}
and so the limit is simply the spatially flat $(k=0)$ FLRW spacetime.

\subsection{Circular orbits in spherical symmetry}\index{circular orbits}

If we have spherical symmetry, the coordinate system adapted to a circling observer will have a time independent $g_{ij}$, as throughout the path of the geodesic there is no variation in the gravitational field. 

Now taking the limit in the appropriate manner, we remove the dependence on the spatial variables. Since our final Bianchi type I spacetime has no time dependence, we can simply rescale coordinates to get the standard form for the Minkowski spacetime. 


\subsection{Radial infall for the Schwarzschild black hole}\index{Schwarzschild spacetime}\index{radial infall}

To take the ultra-local limit in the Schwarzschild solution, we first need the metric in synchronous form\index{synchronous coordinates}. Let us start with the standard/Schwarzschild form:
\begin{equation}
\d s^2 = -\left(1-\frac{2M}{r}\right)\,\d t^2 + \frac{1}{1-\frac{2M}{r}}\d r^2 +r^2 \left(\d\theta^2+\sin^2(\theta)\,\d\phi^2\right).
\end{equation}
Now, closely following both section 102 in \cite{Landau}, and \cite{Lopez}, perform a coordinate transform of the form\index{coordinate transform}
\begin{equation}
t - \rho = \int \frac{r}{\left(1-\frac{2M}{r}\right)\sqrt{2Mr}} \d r,
\end{equation}
which defines $\rho$. Furthermore define a new time coordinate, $\tau$, by
\begin{equation}
\tau - \rho = \int \frac{r}{\left(2Mr\right)^{\frac{1}{2}}} \d r.
\end{equation}
So that 
\begin{equation}
\d r = \frac{\sqrt{2Mr}}{r}\left(\d \tau - \d \rho\,\right),
\end{equation}
\begin{equation}\label{schwarzr}
r(\tau, \rho) = \left(\frac{3}{2}\left(\tau - \rho\, \right)\right)^{\frac{2}{3}}\left(2M\right)^{\frac{1}{3}}.
\end{equation}
We see this is synchronous as
\begin{equation}
-\left(1-\frac{2M}{r}\right)\,\d t^2 + \frac{1}{1-\frac{2M}{r}}\,\d r^2 = -\d\tau^2+ \frac{2M}{r(\tau, \rho)}\,\d \rho^2.
\end{equation}
So the Schwarzschild metric in synchronous coordinates is
\begin{equation}
\d s^2 = -\d\tau^2+ \frac{2M}{r(\tau, \rho)}\d \rho^2+r(\tau, \rho)^2 \left(\d\theta^2+\sin^2(\theta)\d\phi^2\right),
\end{equation}
where $r(\tau, \rho)$ is defined by equation (\ref{schwarzr}). These coordinates are known as Lemaitre or Novikov coordinates \cite{Landau, MTW, novikov}\index{Novikov coordinates}\index{Lemaitre coordinates}.

Finally, before taking the limit, shift coordinates
\begin{equation}
\theta'=\theta - \frac{\pi}{2},
\end{equation}
so that
\begin{equation}
\d s^2 = -\d t^2 + \frac{2M}{r(\tau, \rho)} \,\d \rho^2+ r(\tau, \rho)^2\left(\d \theta'^2+\cos^2(\theta')\,\d \phi^2\right).
\end{equation}
Now to take the limit we take $\rho \to 0$, $\theta' \to 0$, such that
\begin{equation}
r(\tau, \rho) \to r(\tau) = \left[\frac{2}{3}\left(\tau\right)\right]^{\frac{2}{3}}(2M)^{\frac{1}{3}},
\end{equation}
and
\begin{equation}
\cos(\theta') \to 1.
\end{equation}
Resulting in
\begin{equation}
\d s^2 = -\d \tau^2 + \frac{2M}{r(\tau)} \,\d \rho^2+ r(\tau)^2\,\left(\d \theta'^2+\d \phi^2\right),
\end{equation}
or explicitly,
\begin{equation}
\d s^2 = -\d \tau^2 +\left(\frac{3M}{\tau}\right)^{\frac{2}{3}} \d \rho^2+ (2M)^{\frac{2}{3}}\left(\frac{3}{2}\tau\right)^{\frac{4}{3}}\left(\d \theta'^2+\d \phi^2\right).
\end{equation}
Now in this form we can change the coordinates again to put into a more familiar form by rescaling the spatial coordinates:
\begin{equation}\label{kasner122}
\d s^2 = -\d \tau^2 +\tau^{-\frac{2}{3}} \d \rho^2+ \tau^{\frac{4}{3}}\left(\d \theta'^2+\d \phi^2\right),
\end{equation}
which is the Kasner solution with exponents $\left(-\frac{1}{3}, \frac{2}{3}, \frac{2}{3}\right)$.

\vspace{20pt}
\noindent
Equation (\ref{kasner122}) is a type of Kasner solution (equation (\ref{kasnerdef})), and this particular Kasner solution is also known as the $AIII$ metric \cite{Ehlers, Podolskyexact}, and was initially investigated as a variation on the Schwarzschild metric:
\begin{equation}
\d s^2 = \frac{2M}{r}\,\d t^2 - \frac{r}{2M}\,\d r^2 +r^2 \left(\d\theta^2+\sin^2(\theta)\,\d\phi^2\right).
\end{equation}
Then the metric (\ref{kasner122}) can be obtained by performing the coordinate transformation
\begin{equation}
t = \left(\frac{3}{4M}\right)^{\frac{1}{3}}\rho; \qquad r= \left(\frac{3M}{2}\right)^{\frac{1}{3}}\tau^{\frac{2}{3}}.
\end{equation}

\section{Discussion}

We have developed an ultra-local limit that can be applied to any timelike geodesic of an arbitrary spacetime. This construction seems to be the closest timelike analogue to the lightlike Penrose limit. Although the construction of this ultra-local limit is similar to the Penrose limit, it does not share all the familiar properties of that limit, and in particular the hereditary property does not follow.

The result of this limit is a general Bianchi type I metric, which has been studied thoroughly in cosmology, both as a simple cosmological model and as an ultra-local approximation to more general scenarios. In the context of this limit, in contrast to these cosmological settings, there is no clear reason to restrict to diagonal metrics, and we have outlined clearly when this is a reasonable assumption. Keeping this caveat in mind, the vast amount of literature regarding Bianchi type I spacetimes can be a resource of knowledge to apply to the ultra-local limit developed herein.

Working in this general framework we have explored the classical properties of the Bianchi type I spacetime, including an examination of Einstein equations, covariant conservation, and the Raychaudhuri equation. Furthermore we have worked through a few simple examples of this limit.

\chapter{Conclusions}

This thesis has dealt with two research problems in classical general relativity, both of which have employed and extended our knowledge of simple, but physically useful, exact solutions. These two research problems can be regarded independently, and as such the conclusions at the end of each chapter serve to summarise the results for each problem. However, the overall key conclusions and some possible extensions are stated below.

\vspace{28pt}

\noindent
For the first research problem, strong-field gravitational waves were investigated. Ways of representing polarisation modes for these waves were examined, particularly the previously unaddressed issue of how to write down general polarisation modes for the Rosen form. Two forms are commonly used for gravitational waves. In the Brinkmann form it is entirely trivial to construct any polarisation, whereas it was previously unclear how to do so for the Rosen form. 

It was discovered that it was possible to decouple the vacuum equations for the Rosen form into parts depending only on an arbitrary walk in some appropriately defined polarisation space, and a further differential equation relating this walk to the ``envelope'' function. In (3+1) dimensions this polarisation space is the hyperbolic plane. 
From this it was possible to write down a simple algorithm for constructing polarisation modes. This was furthermore extended to arbitrary dimensions. The particularly important, previously unknown, circular polarisation was explicitly written down. 

This investigation essentially closes a long-standing gap in our understanding of the Rosen form and strong-field polarisation. An addition that could be made is the construction of further specific polarisations, including an explicit construction of the elliptical polarisation.

\vspace{28pt}

\noindent
The second research problem was somewhat more open-ended. An ultra-local limit was developed as a close timelike analogue the well-known and well studied lightlike Penrose limit. Although this limit seems to be the closest timelike analogue to the Penrose limit, this limit does not share all the familiar properties of lightlike Penrose limits.

The result of this ultra-local limit is a general Bianchi type I spacetime. Such spacetimes have been most frequently been studied both as simple cosmological models and in approximations to more general cosmological scenarios, where gradient terms may become negligible. In these cosmological contexts it is most common to examine diagonal metrics with vacuum, dust, or perfect fluid scenarios. Such scenarios do not cover the full range of generality desired for our limit, as we saw from an examination of the effect the limit had on various scenarios for matter content. Thus we have explicitly set forward the necessary and sufficient conditions under which a Bianchi type I spacetime is diagonal. 

We have examined several aspects of the Bianchi type I spacetime keeping full generality. This includes various efforts to write down the useful geometrical tensors for the Bianchi type I spacetime as well as an examination and interpretation of the Einstein equations, covariant conservation, and Raychaudhuri equation. Furthermore, we have looked at a few simple examples of how this limit is approached in various scenarios.

The work on this problem does raise some further avenues for investigation. Penrose limits have generated considerable interest, and have been very well studied in a number of specific spacetimes. We have already worked through a few simple examples of this limit, and this work could probably be usefully extended, hopefully finding a small number of limits for various physically relevant spacetimes. The question still remains about exactly how much useful information is retained by this ultra-local limit, a question that has been extensively addressed in the case Penrose limits over many years. One goal here would be to completely characterise at least some aspects of the original spacetimes by properties of the various Bianchi type I spacetimes that result from the limits of such spacetimes.

\appendix

\titleformat{\chapter}[frame]{\normalfont}{\filright APPENDIX \thechapter}{8pt}{\Huge\rm\filcenter}

\titleformat{\section} {\titlerule\vspace{-3.8ex}\normalfont\bf} {\thesection.}{.5em}{}

\titlespacing*{\section}{0pc}{*8}{*5}

\chapter{Publications}

\vspace*{55pt}

\section{General polarization modes for the Rosen gravitational wave}

\begin{center}
Bethan Cropp and Matt Visser
\end{center}

\noindent
Strong-field gravitational plane waves are often represented in either the Rosen or Brinkmann forms. While these two metric ansatz\"e are related by a coordinate transformation, so that they should describe essentially the same physics, they rather puzzlingly seem to treat polarization states quite differently. Both ansatz\"e deal equally well with $+$ and $\times$ linear polarizations, but there is a qualitative difference in the way they deal with circular, elliptic, and more general polarization states. In this article we will develop a general formalism for dealing with arbitrary polarization states in the Rosen form of the gravitational wave metric, representing an arbitrary polarization by a trajectory in a suitably defined two dimensional hyperbolic plane. 

\begin{center}
Published: Classical and Quantum Gravity {\bf 27} (2010) 165022.

\noindent
arXiv: gr-qc/1004.2734
\end{center}

\newpage

\vspace*{100pt}

\section{Polarization modes for strong-field gravitational waves}

\begin{center}
Bethan Cropp and Matt Visser
\end{center}

\noindent
Strong-field gravitational plane waves are often represented in either the Rosen or Brinkmann forms. These forms are related by a coordinate transformation, so they should describe essentially the same physics, but the two forms treat polarization states quite differently. Both deal well with linear polarizations, but there is a qualitative difference in the way they deal with circular, elliptic, and more general polarization states. In this article we will describe a general algorithm for constructing arbitrary polarization states in the Rosen form.

\begin{center}
To be published in the Journal of Physics: Conference Series.\\
Proceedings of the Spanish Relativity Meeting (ERE2010).

\noindent
arXiv: gr-qc/1011.5904
\end{center}

\newpage

\vspace*{100pt}

\section{Any spacetime has a Bianchi type~I spacetime as a limit}

\begin{center}
Bethan Cropp and Matt Visser
\end{center}

\noindent
Pick an arbitrary timelike geodesic in an arbitrary spacetime. We demonstrate that there is a particular limiting process, an ``ultra-local limit'', in which the immediate neighbourhood of the timelike geodesic can be ``bl\-own up'' to yield a general (typically non-diagonal) Bianchi type~I spacetime. This process shares some (but definitely not all) of the features of the Penrose limit, whereby the immediate neighbourhood of an arbitrary null geodesic is ``blown up'' to yield a $pp$-wave as a limit.

\begin{center}
Published: Classical and Quantum Gravity {\bf 28} (2011) 055007.

\noindent
arXiv: gr-qc/1008.4639
\end{center}


\bibliographystyle{eprint}

\addcontentsline{toc}{chapter}{Bibliography}
\bibliography{myrefs}

\addcontentsline{toc}{chapter}{Index}
\printindex

\end{document}